\documentclass[10pt,final, journal, twocolumn]{IEEEtran}
\usepackage{threeparttable} 
\usepackage{graphicx}
\usepackage{float}
\usepackage{picinpar}
\usepackage{rotating}
\usepackage{cite}

\usepackage{psfrag}
\usepackage{placeins}
\usepackage{times}
\usepackage{color}
\usepackage{amsmath}
\usepackage{amsfonts}
\usepackage{subfig}

\usepackage{caption}
\usepackage{bm}

\usepackage{pifont}
\usepackage{amssymb}
\usepackage{setspace}
\usepackage{changebar}
\usepackage{flafter}
\DeclareGraphicsExtensions{.eps,.pdf}
\usepackage{booktabs}
\usepackage{tabularx}
\usepackage{array}

\newcommand{\PreserveBackslash}[1]{\let\temp=\\#1\let\\=\temp}
\newcolumntype{C}[1]{>{\PreserveBackslash\centering}p{#1}}
\newcolumntype{R}[1]{>{\PreserveBackslash\raggedleft}p{#1}}
\newcolumntype{L}[1]{>{\PreserveBackslash\raggedright}p{#1}}

\usepackage{mdwmath}
\usepackage{mdwtab}
\usepackage{dblfloatfix}
\usepackage{morefloats}
\usepackage{url}

\usepackage{cite}
\usepackage{circledsteps}

\usepackage{subfig}

\begin{document}
\title{A Cyber–Physical Routing Protocol Exploiting Trajectory Dynamics for Mission-Oriented Flying Ad Hoc Networks}
\author{\IEEEauthorblockN{Die Hu$^{\text{a.b}}$, Shaoshi Yang$^{\text{a,c,*}}$, Min Gong$^\text{b}$, Zhiyong Feng$^{\text{a,c}}$, and Xuejun Zhu$^\text{b}$}\\

\thanks{$^\text{a}$ School of Information and Communication Engineering, Beijing University of Posts and Telecommunications, Beijing, 100876, China}%
\thanks{$^\text{b}$ China Academy of Launch Vehicle Technology, Beijing 100076, China}%
\thanks{$^\text{c}$ Key Laboratory of Universal Wireless Communications, Ministry of Education, Beijing 100876, China.} %
\thanks{$^*$\emph{Corresponding author}: S. Yang (email: shaoshi.yang@bupt.edu.cn).}
}

\maketitle
\begin{abstract}
As a special type of mobile ad hoc network (MANET), the flying ad hoc network (FANET) has the potential to enable a variety of emerging applications in both civilian wireless communications (e.g., 5G and 6G) and the defense industry. The routing protocol plays a pivotal role in FANET. However, when designing the routing protocol for FANET, it is conventionally assumed that the aerial nodes move randomly. This is clearly inappropriate for a mission-oriented FANET (MO-FANET), in which the aerial nodes typically move toward a given destination from given departure point(s), possibly along a roughly deterministic flight path while maintaining a well-established formation, in order to carry out certain missions. In this paper, a novel cyber–physical routing protocol exploiting the particular mobility pattern of an MO-FANET is proposed based on cross-disciplinary integration, which makes full use of the mission-determined trajectory dynamics to construct the time sequence of rejoining and separating, as well as the adjacency matrix for each node, as prior information. Compared with the existing representative routing protocols used in FANETs, our protocol achieves a higher packet-delivery ratio (PDR) at the cost of even lower overhead and lower average end-to-end latency, while maintaining a reasonably moderate and stable network jitter, as demonstrated by extensive ns-3-based simulations assuming realistic configurations in an MO-FANET.
\end{abstract}


\makeatletter
\def\hlinewd#1{%
  \noalign{\ifnum0=`}\fi\hrule \@height #1 \futurelet
   \reserved@a\@xhline}
\makeatother

\IEEEpeerreviewmaketitle

\section{Introduction}
\IEEEPARstart{M}{odern} unmanned aerial vehicles (UAVs) are defining a new paradigm for executing complex tasks, such as emergency aid [1], geological exploration [2], and border surveillance [3]. Compared with an individual UAV equipped with advanced facilities, a formation of multiple UAVs or a larger scale UAV swarm, is able to perform challenging tasks with better effect, faster convergence, and lower cost [4], [5] when empowered by the flying ad hoc network (FANET) technology, which is based on inter-node networking and cooperation. A FANET does not require fixed terrestrial infrastructure, which makes it particularly attractive for a wide range of applications.

The routing protocol plays a pivotal role in a FANET. In general, the existing routing protocols proposed for mobile ad hoc networks (MANETs) and vehicular ad hoc networks (VANETs) [6] do not perform well in FANETs with high mobility and/or highly dynamic topology, although there are many apparent similarities between these networks [7]. To make FANETs more powerful, substantial research efforts have been devoted to routing protocols. For example, it was shown in Refs. [8], [9] that the signaling overhead and route recalculation cost can be reduced by improving the multi-point relay (MPR) selection method of the optimized link state routing (OLSR) protocol [10]. By utilizing a classic packet-delivery technique based on greedy geographic forwarding [11], the packet-delivery ratio (PDR) and the end-to-end latency of the network are improved when using the ad hoc on-demand distance vector (AODV) [12] routing protocol. Despite these improvements, huge overhead is still required to carry out route discovery, which is prohibitive in large-scale networks. Thus, high-performance low-cost routing protocols are essential for FANETs.

The performance of routing protocols in ad hoc networks varies with mobility models [13], and it is a desirable and significant task to combine mobility traces and protocols together for network design [14]. This is a main feature to consider when studying MANETs. In most mobility models, including indoor mobility models such as random-walk [15], random waypoint, and random direction, as well as outdoor mobility models such as Gauss-Markov [16] and the probabilistic version of random-walk, the location and/or velocity of the nodes are assumed to be random variables.

However, for a mission-oriented FANET (MO-FANET) composed of high-value aerial nodes, such as UAVs and helicopters, a random mobility model diverges from reality. In fact, topology control is highly desirable in an MO-FANET, as it brings a variety of benefits. To elaborate further, an airborne trajectory planning (ATP) system can be deployed to control the mobility pattern of each aerial node in the formation through behavior-based, virtual-structure-based, or leader-follower-based methods [17], or by dynamically using an adaptive protocol [18], thereby ensuring that the tasks are completed as planned and avoiding potentially huge financial loss.

ATP is one of the essential function modules of an MO-FANET, since it is beneficial for improving operational efficiency and safety, and for realizing effective inter-node networking and cooperation in a mission. During a flight, the ATP system can adapt heuristic algorithms (e.g., simulated annealing) [19] to generate trajectories that meet the smoothness constraints of velocity and acceleration. This result can then be applied as the mobility model for simulating routing protocols in a more realistic scenario. Furthermore, formation control, which maintains the geometric pattern of the positions of all the UAVs (e.g., using local neighboring information [20]), can provide a stable topology on a timescale that is large enough to perform routing calculations. Thus, by properly exploiting formation control, the design of the routing protocol for an MO-FANET can be simplified.

More importantly, as a cyber–physical system, an MO-FANET is expected not only to deliver information flow in cyberspace, but also to perform certain maneuvering actions in physical space. Therefore, it is not the best strategy to simply use the routing protocols of traditional ad hoc networks in an MO-FANET.

Against the background described above, this paper makes the following novel contributions to the field.

We propose an innovative cyber–physical routing protocol incorporating interdisciplinary ingredients of trajectory dynamics (CPR-TD) as prior information to improve the overall performance of an MO-FANET. To the best of our knowledge, directly exploiting the native trajectory dynamics of multiple UAVs from the application layer as an input—in addition to considerations from the wireless communication perspective—has never before been reported for designing a routing protocol in the open literature. The significance of our effort is that it potentially opens up a new interdisciplinary research direction with the objective of making full use of information from each protocol layer to support the design of a highly efficient routing protocol.

We build up a mathematical model for characterizing a challenging scenario in which multiple UAVs fly collaboratively in different formations from one place to another in order to execute a certain task. This is an important scenario with broad applications in both civilian and defense industries. More specifically, we model the flying process of the MO-FANET as five phases, which are described by multiple coordinate frames and different sophisticated maneuver actions. In addition, the timescale difference of the physical-space trajectory dynamics and the cyberspace wireless medium-access mechanism is taken into account, which ensures that the topology of the MO-FANET is sufficiently predictable and stable for the operation of the routing protocol.

In comparison with other state-of-the-art protocols, the proposed CPR-TD protocol not only achieves the highest PDR performance, but also attains the highest overhead efficiency (OE) in the context of an MO-FANET. Moreover, in most of the considered cases, it exhibits a lower average end-to-end latency than the benchmarking protocols, as well as a reasonably low and stable network jitter. Our CPR-TD protocol can be implemented with the aid of the application-layer ATP system of individual UAVs. Therefore, it is more suitable for an MO-FANET than the traditional routing protocols used in FANETs.

The rest of this paper is organized as follows. In Section~2, we present the modeling of the flying process of an MO-FANET by considering trajectory dynamics that are described by multiple coordinate frames and sophisticated maneuver actions. In Section~3, we describe our CPR-TD protocol in detail. Extensive ns-3-based simulations that assume realistic network configurations, along with a detailed discussion, are presented in Section~4. Finally, we offer our conclusions in Section~5.   

\section{Modeling of the flying process of an MO-FANET}
\subsection{Description of the application scenario}
In an emergency or disaster rescue scenario, a terrestrial strategy is often inadequate to control the situation effectively, and a FANET-based strategy becomes indispensable to the execution of the mission. For example, in the recent Australian bushfires, Amazon rainforest fires, and California wildfires with a global impact, firefighting UAVs have played an important role; however, they are typically operated in an individual manner.

\begin{figure}[t]
\centering
\includegraphics[width=0.45\textwidth]{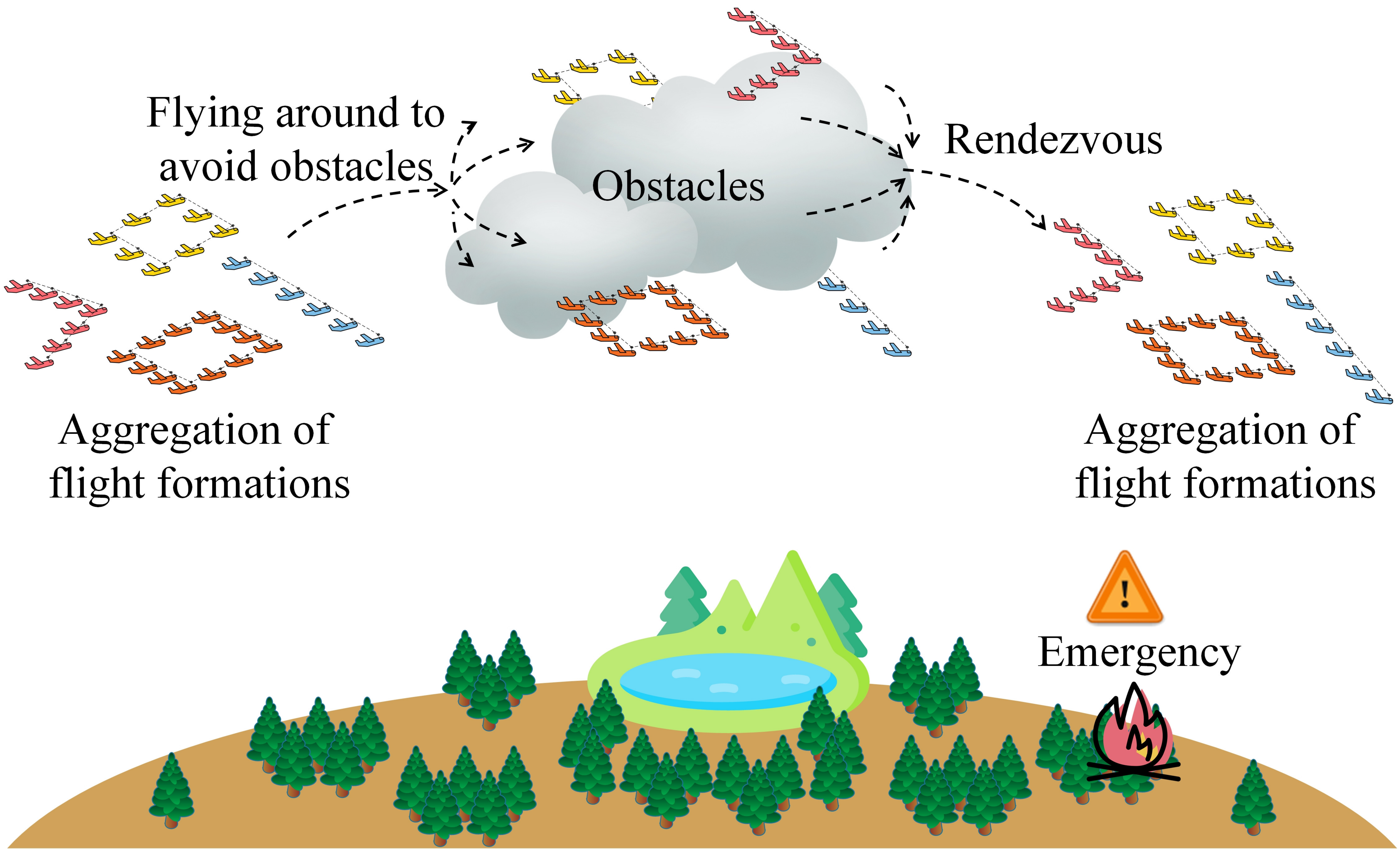}
\caption{An application example of the flight formation in an MO-FANET.}
\label{fig:application}
\end{figure}

We consider a challenging scenario in which multiple UAVs fly collaboratively in different formations from one place to another in order to execute a certain task, as shown in Fig.\ref{fig:application}, with four flight formations. In order to guarantee flight security and task execution efficiency, the individual ATP systems collaboratively find the optimal trajectory for each UAV by means of distributed optimization algorithms and maintain the formation of each UAV cluster. To this end, it is necessary for the UAVs to capture the motion information of other UAVs in real time and then perform trajectory calculations. During the flight, the formations may maneuver, such as by flying around, in response to obstacles or hazardous situations. Thus, the dynamic motion state of the MO-FANET can be classified into four states: \Circled{1} flight formation aggregation;  \Circled{2} flying around;  \Circled{3} rendezvous, which are followed by  \Circled{4} a return to flight formation aggregation.

For a flight formation, it is worth noting that, according to the results of ATP, the state of motion can be predicted to some degree within a certain time period, which is dramatically different from the random mobility model often used in the open literature.

\subsection{Trajectory generator}
\begin{figure}[t]
\centering
\includegraphics[width=0.45\textwidth]{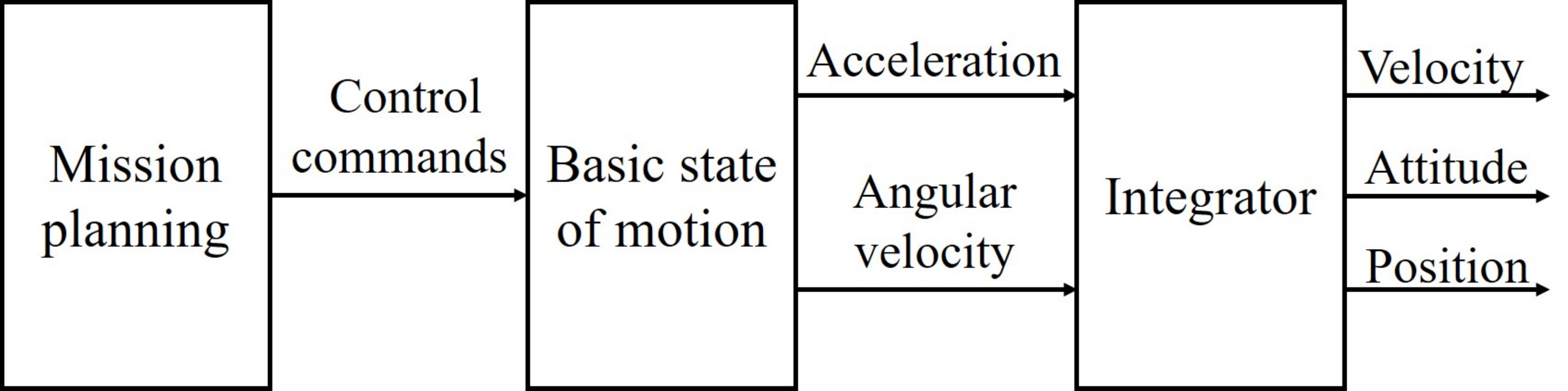}
\caption{The method of trajectory parameter acquisition.}
\label{fig:trajectory}
\end{figure}

For a clearer description of the flying process of the MO-FANET considered here, we adopt the trajectory generator of the strapdown inertial navigation system (SINS). This generator can reproduce the UAV trajectory as realistically as possible, and has been widely used in analyses and experiments on the strapdown inertial navigation algorithm and the integrated navigation algorithm in the area of flight control [21]. In this paper, we focus on acquiring the trajectory parameters of UAVs; hence, the specific force and inertial tensor will not be considered in our routing protocol design. The flowchart in Fig. \ref{fig:trajectory} summarizes the method of trajectory parameter acquisition. To facilitate understanding, in what follows, we first introduce the definitions of related coordinate frames. We then describe typical maneuver actions and finally present our mobility model.

\subsubsection{Definitions of coordinate frames}
\begin{figure}[ht]
\centering
\includegraphics[width=0.4\textwidth]{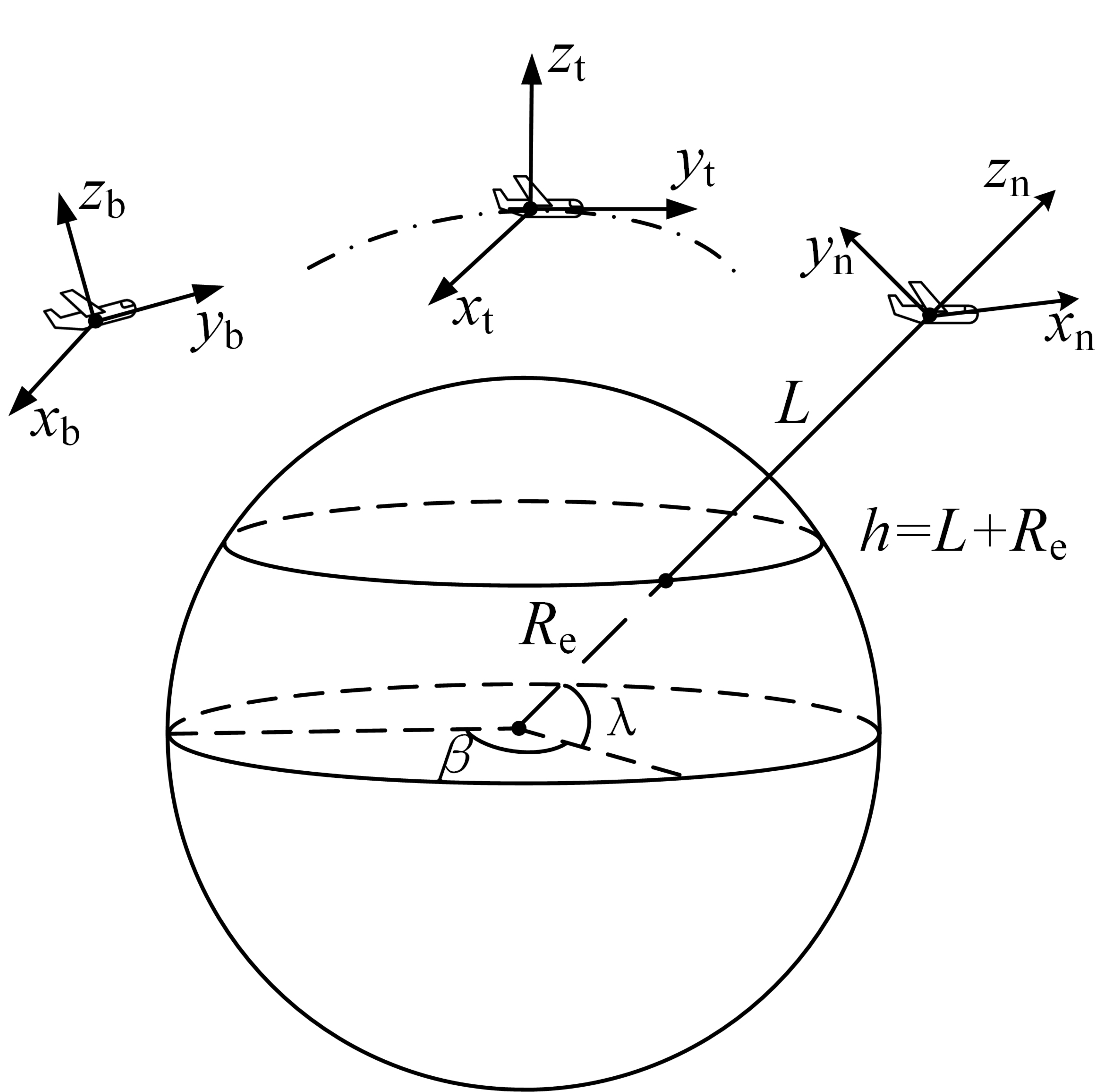}
\caption{ The coordinate frames. $\beta$: latitude;  $\lambda $: longitude; $h$: altitude; $R_{\textrm{e}}$: equatorial radius of the earth; $L$: the length from the aircraft to the point of intersection on the surface of the earth; ${x_{\textrm{b}}}$, ${y_{\textrm{b}}}$, ${z_{\textrm{b}}}$: directions of body-axis coordinate frame; ${x_{\textrm{t}}}$, ${y_{\textrm{t}}}$, ${z_{\textrm{t}}}$: directions of trajectory coordinate frame; ${x_{\textrm{n}}}$, ${y_{\textrm{n}}}$, ${z_{\textrm{n}}}$: directions of navigation coordinate frame.}
\label{fig:Axis}
\end{figure}

The related coordinate systems involved in this paper are shown in Fig. \ref{fig:Axis} and are briefly introduced below.

\begin{itemize}
\item[(1)]  \textbf{Body-axis coordinate frame (b-frame):} The $y$-axis points forward through the nose of the aircraft and is denoted as $y_{\textrm{b}}$; the $x$-axis points to the right of the $y$-axis (facing the pilot’s direction of view), perpendicular to the $y$-axis, and is denoted as $x_{\textrm{b}}$; and the $z$-axis points up through the bottom of the aircraft, perpendicular to the $xy$-plane, and satisfies the right-hand rule.
\item[(2)]  \textbf{Terrestrial coordinate frame (e-frame):} The e-frame uses a triplet (latitude, longitude, and altitude) to denote the location of an object; for example, in Fig. \ref{fig:Axis}, the e-frame defines $\beta$ as the latitude, $\lambda $  as the longitude, $h$ as the altitude, $R_{\textrm{e}}$ as the equatorial radius of the earth, and $L$ as the length from the aircraft to the point of intersection on the surface of the earth.
\item[(3)]  \textbf{Navigation coordinate frame (n-frame):} The n-frame defines  ${x_{\textrm{n}}}$, ${y_{\textrm{n}}}$ and ${z_{\textrm{n}}}$  as the three directions of east, north, and up, respectively.
\item[(4)] \textbf{Trajectory coordinate frame (t-frame):} The t-frame defines the horizontal right as ${x_{\textrm{t}}}$ and the direction of motion that is tangent to the trajectory as ${y_{\textrm{t}}}$. Again, ${z_{\textrm{t}}}$ is defined according to the right-hand rule.
\end{itemize}

The attitude angles of the UAVs, which reflect the flight conditions, are shown in Fig. \ref{fig:Angle}, and their definitions are given below. $\psi $ denotes the course angle, which is the angle between the geographic North Pole and the horizontal projection of the longitudinal axis of the aircraft body; $\theta $  denotes the pitch angle, which is the angle between the horizontal projection of the longitudinal axis of the aircraft body and this longitudinal axis itself; $\gamma $ denotes the roll angle, which is the angle between the vertical axis and the vertical plane.

We assume that $\psi $, $\theta$ and $\gamma$ are linear functions of time $t$, which is typically realistic.
 
\begin{figure}[tbp]
\centering
\includegraphics[width=0.35\textwidth]{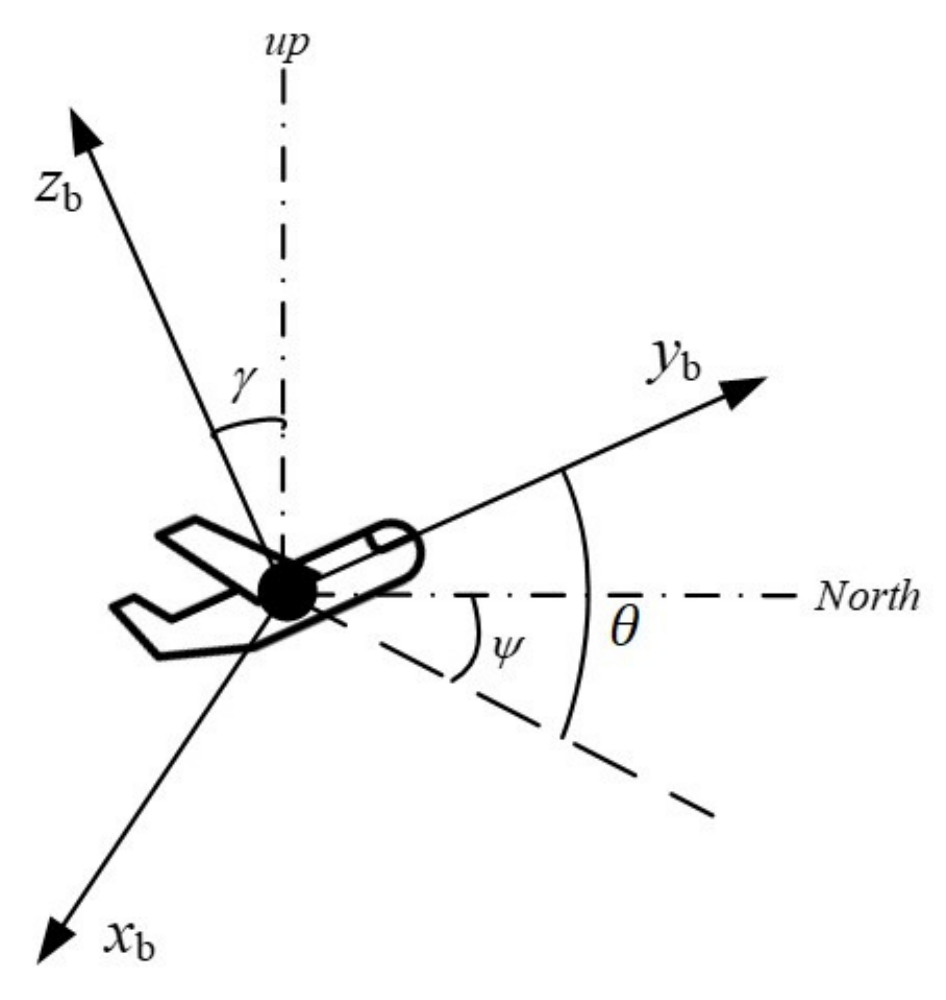}
\caption{ The attitude angles of an aircraft.}
\label{fig:Angle}
\end{figure}

Through revolution and translational movement, we can convert coordinates between different coordinate frames [22]. To be specific,  $[x_{\textrm{n}}, y_{\textrm{n}}, z_{\textrm{n}}]^T = {\boldsymbol C}_{\textrm{t}}^{\textrm{n}} [x_{\textrm{t}}, y_{\textrm{t}}, z_{\textrm{t}}]^T$, where T represents the transpose operator; ${\boldsymbol C}_{\textrm{t}}^{\textrm{n}}$ is the rotation array from the t-frame to the n-frame and is given as follows:
\begin{equation}
\label{C_nt}
{\boldsymbol C}_{\textrm{t}}^{\textrm{n}} = \left[ {\begin{array}{*{20}{c}}
{\cos \psi }&{\cos \theta \sin \psi }&{ - \sin \theta \sin \psi }\\
{ - \sin \psi }&{\cos \theta \cos \psi }&{ - \sin \theta \cos \psi }\\
0&{\sin \theta }&{\cos \theta }
\end{array}} \right].
\end{equation}

In addition, given the influence of the oblateness of the earth, the change of position in the e-frame can be obtained from the change of motion state in the t-frame by means of the following differential equations:
\begin{equation}
\label{diff}
\left\{ {\begin{array}{*{20}{l}}
{\left[ {\frac{{\partial \theta }}{{\partial t}},\frac{{\partial \gamma }}{{\partial t}},\frac{{\partial \psi }}{{\partial t}}} \right] = {{\boldsymbol{\omega }}_{\rm{b}}},}\vspace{1ex}\\
{\frac{{\partial {{\boldsymbol {v}}_{\rm{n}}}}}{{\partial t}} = {{\boldsymbol {a}}_{\rm{n}}},}\vspace{1ex}\\
{\frac{{\partial \beta }}{{\partial t}} = {{\bm v}_{y,{\rm{n}}}}/({R_M} + L),}\vspace{1ex}\\
{\frac{{\partial \lambda }}{{\partial t}} = {{\bm v}_{x,{\rm{n}}}}\sec \beta /({R_N} + L),}\vspace{1ex}\\
{\frac{{\partial h}}{{\partial t}} = {{\bm v}_{z,{\rm{n}}}}},
\end{array}} \right.
\end{equation}
where ${\boldsymbol\omega}_{\textrm{b}}$ is the attitude angular velocity vector in the b-frame, ${\bm v}_{\textrm{n}}$ is the velocity vector in the n-frame, ${\boldsymbol a}_{\textrm{n}}$ is the acceleration vector in the n-frame, ${R_N} = {R_e}(1 - 2e + 3e\sin 2\beta)$ is the radius of curvature in the prime vertical, and ${R_M} = {R_e}(1 + e\sin 2\beta)$ is the radius of the curvature in the meridian, with the constant $e$ representing the oblateness of the earth. It should be noted that, in Eq. \eqref{diff}, ${\boldsymbol a}_{\textrm{n}}$ is obtained by changing the acceleration from the t-frame to the n-frame—that is, by translation of the coordinate axis using

\begin{equation}
\label{eq:t_to_n}
{{\boldsymbol a}_{\textrm{n}}} = {\boldsymbol C}_{\textrm{t}}^{\textrm{n}}{{\boldsymbol a}_{\textrm{t}}},
\end{equation}
where ${\boldsymbol a}_{\textrm{t}}$ is the acceleration vector in the t-frame.

We then obtain the trajectory, velocity, and attitude of the aircraft by solving the above differential equations using the fourth-order Runge–Kutta method [23]. In fact, any other feasible integral method can also be used to attain the same results.

\subsubsection{ Description of typical maneuver actions}
A change in the motion state of a UAV is caused by the attitude angular velocity in the b-frame and the acceleration in the t-frame. The typical maneuver actions are mathematically characterized as follows.
\begin{itemize}
\item[(1)] \textbf{Uniform rectilinear movement.} In this state, neither the attitude angular velocity nor the acceleration is changed; thus, we have
\begin{align}
{{\boldsymbol \omega} _{\textrm b}} & = [0,0,0],\\
{{\boldsymbol a}_{\textrm t}}  & = [0,0,0].
\end{align}
\item[(2)] \textbf{Uniform acceleration or deceleration.} In this state, only the acceleration is changed, and the UAV continues to accelerate or decelerate forward with the acceleration . Thus, we have
\begin{align}
{{\boldsymbol \omega} _{\rm{b}}} & =  [0,0,0],\\
{{\boldsymbol a}_{\textrm{t}}}  & =  [0,a,0].
\end{align}
\item[(3)] \textbf{Coordinated turn.} The UAV turns by banking according to the practical conditions; hence, this state is divided into three phases: banking, turning, and the transition from turning to level flight. In the first phase, the UAV’s roll angle is changed from zero to $\gamma $ with a constant angular velocity of $\frac{\partial \gamma}{\partial t}$ in order to prepare for turning. Thus, we have
\begin{align}
{{\boldsymbol \omega} _{\rm{b}}} & = {\Big [}0,\frac{\partial \gamma}{\partial t},0 {\Big ]},\\
{{\boldsymbol a}_{\textrm{t}}} & = [0,0,0].
\end{align}

In the second phase, the UAV maintains a roll angle equal to $\gamma $  and starts turning with a constant angular velocity of $\frac{\partial \psi}{\partial t}  $; the centripetal force is provided by the lift generated by banking. Then, we have
\begin{align}
{{\boldsymbol \omega} _{\rm{b}}} & = {\Big [}0,0,\frac{\partial \psi}{\partial t} {\Big ]},\\
{{\boldsymbol a}_{\textrm{t}}} & = [0,0,g\tan \gamma ],
\end{align}
where $g$ is the acceleration of gravity.
  
In the third phase, the UAV levels off after turning and changes the roll angle with a constant angular velocity of $ - \frac{\partial \gamma}{\partial t}  $. Thus we have 
\begin{align}
{{\boldsymbol \omega} _{\rm{b}}} & = {\Big [}0, - \frac{\partial \gamma}{\partial t}  ,0 {\Big ]}, \\
{{\boldsymbol a}_{\textrm{t}}} &  = [0,0,0]. 
\end{align}

\item[(4)]  \textbf{Climb or descend.} This state can also be divided into three phases. In the first phase, the UAV performs a circular motion with radius $r$ on the vertical plane and changes the pitch angle from zero to $\theta $ with a constant angular velocity $\frac{\partial \theta}{\partial t}$ of in order to prepare for climbing or descending. Then, we have
\begin{align}
{{\boldsymbol \omega} _{\rm{b}}} & = {\Big [}\frac{\partial \theta}{\partial t} ,0,0,{\Big ]}\\
{{\boldsymbol a}_{\textrm{t}}}  & = {\Big [} 0,0,{{\Big(}{\frac{\partial \theta}{\partial t}  }{\Big )}^2}r {\Big ]}. 
\end{align}
In the second phase, the UAV maintains the pitch angle and carries out a uniform rectilinear movement to climb or descend. Hence, we have
\begin{align}
{{\boldsymbol \omega} _{\rm{b}}} & = [0,0,0],\\
{{\boldsymbol a}_{\textrm{t}}} & = [0,0,0].
\end{align}
The third phase is the inverse process of the first phase, and the UAV levels off after climbing or descending. Thus, we have
\begin{align}
{{\boldsymbol \omega} _{\rm{b}}} & = {\Big [} - \frac{\partial \theta}{\partial t}  ,0,0 {\Big ]},\\
{{\boldsymbol a}_{\textrm{t}}} & = {\Big [} 0,0, - {{\Big (}{\frac{\partial \theta}{\partial t}} {\Big )}^2}r {\Big ]}. 
\end{align}
\end{itemize}

\begin{figure}[tbp]
\centering
\includegraphics[width=0.45\textwidth]{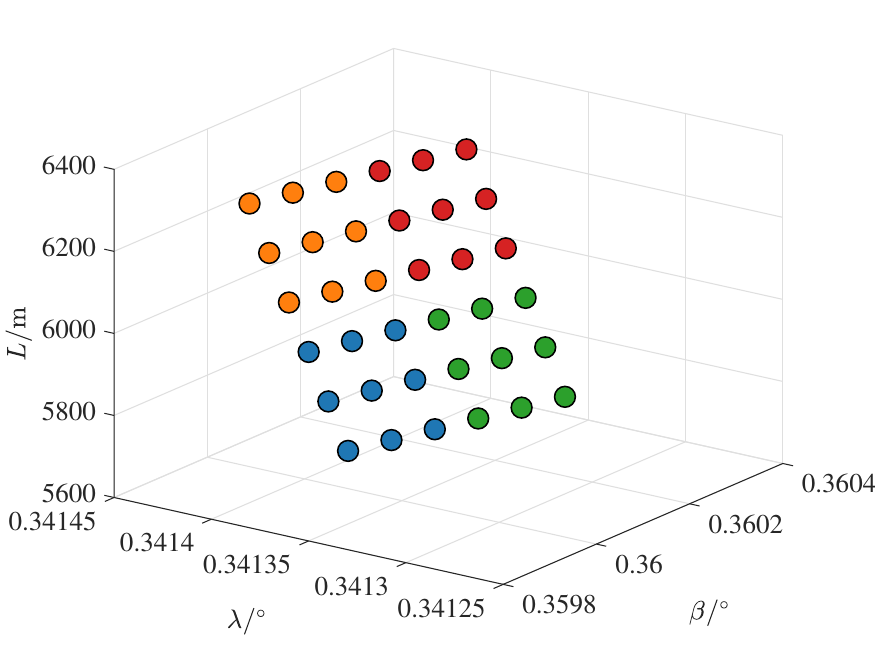}
\caption{The flight formation.}
\label{fig:Flight_formation}
\end{figure}
\subsubsection{Mobility model}
To perform a more realistic study, we consider a scenario in which the UAVs can avoid obstacles during the flight through collaboration. As shown in Fig. \ref{fig:Flight_formation}, we assume that all the UAVs are initially split into four groups, which are represented by four different colors. For simplicity, the UAVs in each group are assumed to be relatively static and maintain a diamond shape by using the airborne formation control system during the whole flight process.
\begin{figure}[tbp]
\vspace{-0.4cm}
\centering
\includegraphics[width=0.45\textwidth]{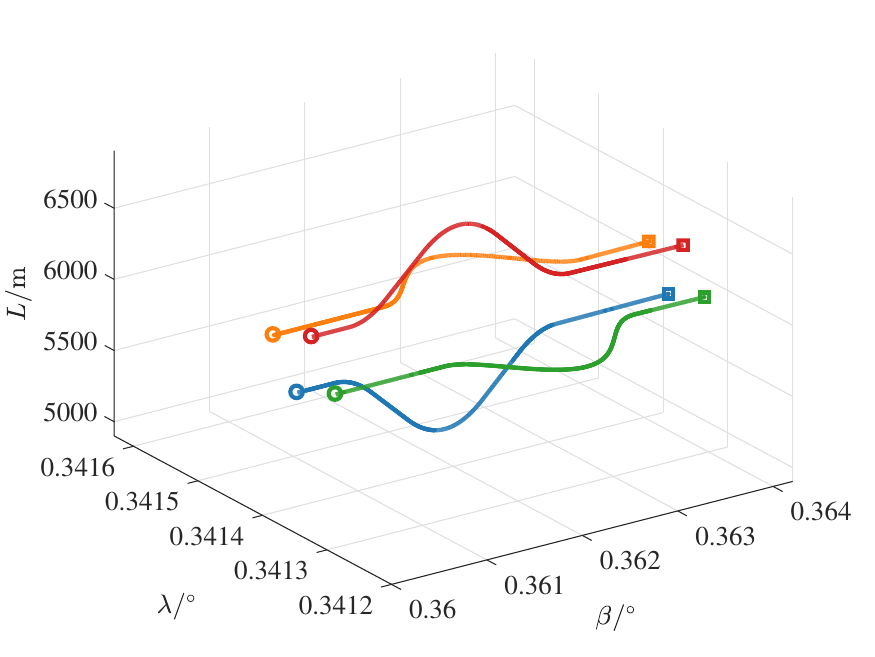}
\caption{The trajectories of the four UAV groups.}
\label{fig:simulation_trajectory}
\end{figure} 

The formation or maneuver actions can be changed in light of the actual conditions. When obstacles are detected, the UAV groups first carry out separation and then recombination to avoid obstacles. For clarity, the trajectories of all the UAV groups are shown in Fig. \ref{fig:simulation_trajectory}, where each trajectory with a particular color corresponds to the group of the same color in Fig. \ref{fig:Flight_formation}.

\section{The proposed cyber–physical routing protocol exploiting the trajectory dynamics of an MO-FANET}
In this section, we present the details of the proposed CPR-TD of the MO-FANET. Our protocol is the result of interdisciplinary study, where the output of the ATP system is invoked to facilitate the route selection.
\subsection{Packet header}
\begin{figure}[ht]
\centering
\includegraphics[width=2.5in]{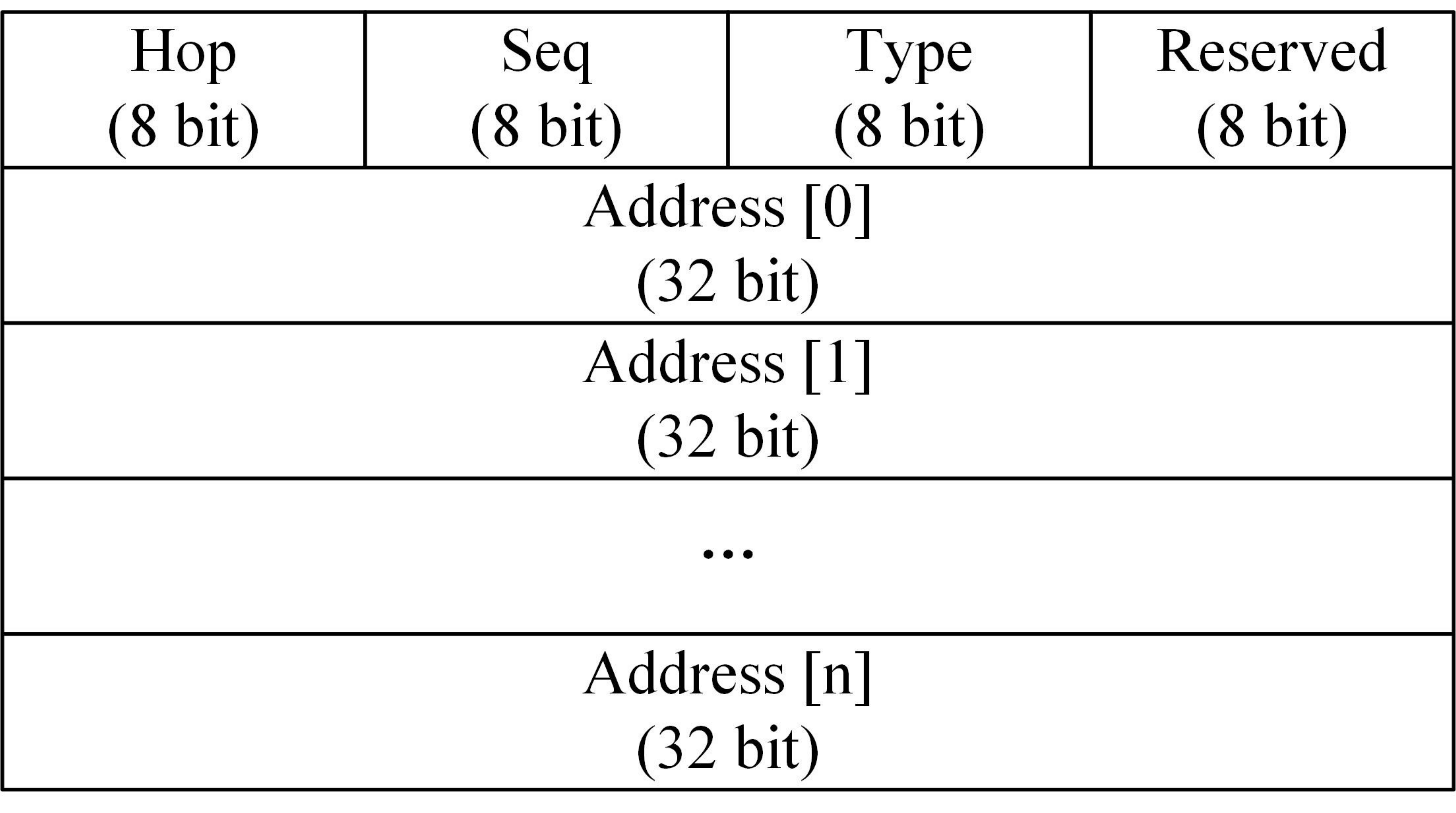}
\caption{The packet header format.}
\label{fig:header}
\end{figure}

Inspired by the Control packet in the dynamic source routing (DSR) protocol, the routing information in our protocol is stored in the packet header, which constitutes part of the Data packet and comprises the Control packet. The packet header is designed to inform other aerial nodes of the number of hops on the route, to uniquely index each packet, to identify the packet type, and to indicate the packet-delivery route, as shown in Fig. \ref{fig:header}. We assume that the same ATP system is individually deployed on all the nodes of the MO-FANET, while both signaling and action synchronizations can be achieved between the distributed nodes. We also assume that the MO-FANET is in a steady state in each session, and that it can change to a different state in another session based on the periodical state broadcast from each aerial node involved. Then, in each session, the information source nodes individually calculate the route according to the output of the ATP system and the available communication resources, and store the number of hops in the “Hop” field and the address of the mth node on the route in the “Address [m]” field. In addition, whenever a packet is generated by a given source node, the packet is assigned an 8-bit sequence number, which, in conjunction with the “Address [0]” field (i.e., the address of the source node) and the “Address [n]” field (i.e., the address of the destination node), characterizes the one-to-one correspondence between the packet and the source-destination pair, and is stored in the “Seq” field. Furthermore, the “Type” field in the packet header stores the packet type, including the Data packet, the route reply (RREP) packet, and the route error (RRER) packet. It should be noted that the RREP and RRER packets are regarded as the Control packets. Finally, we reserve 8 bits for functions that are yet to be defined.

\subsection{Route establishment}
Unlike the existing schemes, our protocol has no route request (RREQ) process for establishing a route from the source to the destination. As a result, the overhead is significantly reduced. In fact, the source node in our protocol exploits the information on the trajectory dynamics from certain applications to establish the route. From a cyberspace perspective, information on the mission and trajectory planning is shared among the UAVs of the MO-FANET. The routing protocol that operates on the network layer can obtain the location information of each node from the ATP system via the application layer, or from the physical layer signal-processing module (e.g., the global positioning system (GPS) module or the in-network distributed collaborative positioning module) in order to support the calculation of routes. From the perspective of physical space, our protocol can provide more reliable data delivery, which is required by the ATP system to control the trajectories of the UAVs. More specifically, we use the notification of the rejoining and separating time between all the aerial nodes to obtain the time list of rejoining and separating, and use the output of the formation configuration to obtain the adjacency matrix. A flowchart is given in Fig. 8 to clarify the route-establishing process in our packet-delivery scheme.
\begin{figure*}[htbp]
\begin{center}
\includegraphics[angle = 0,width = 15cm]{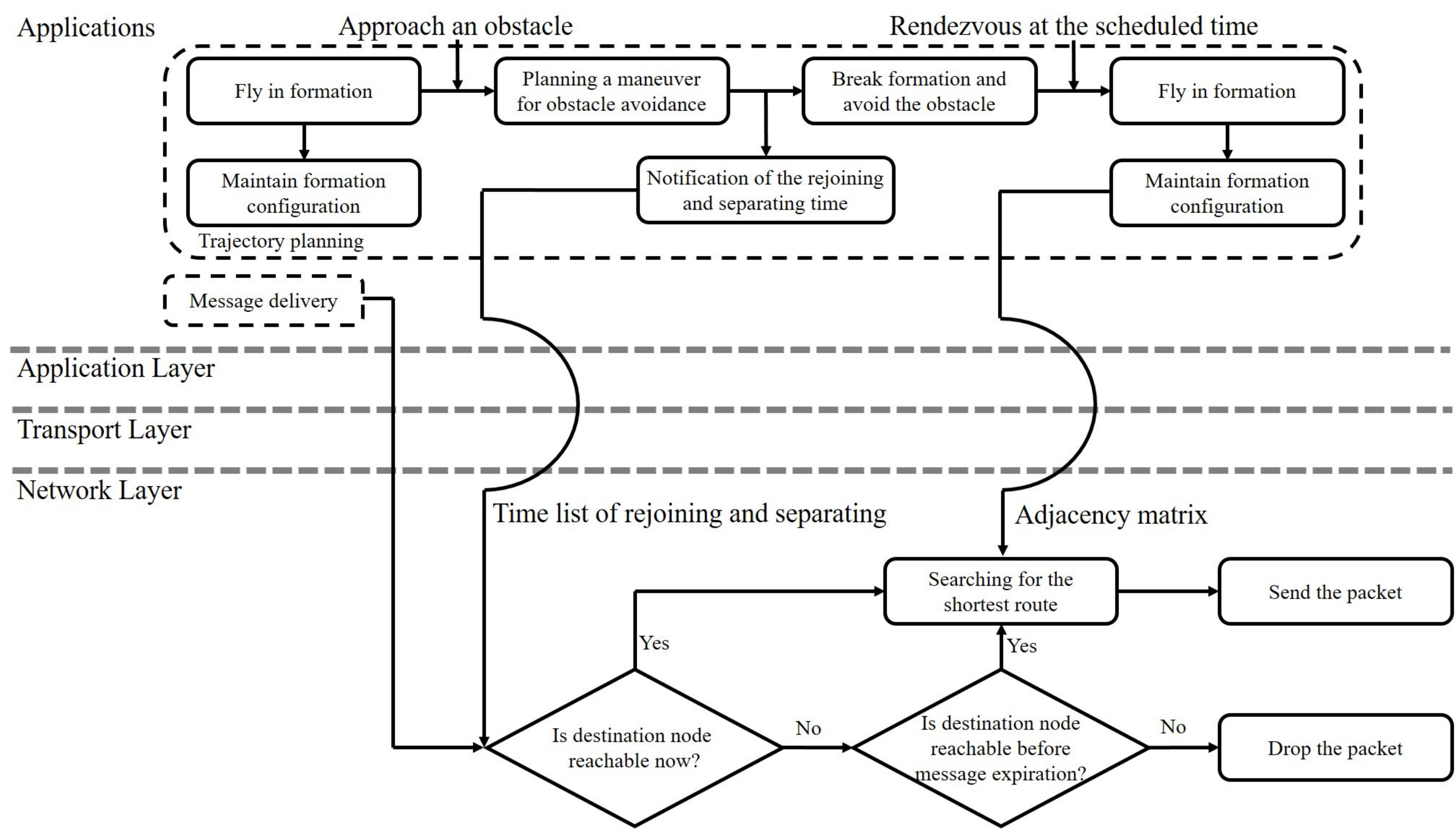}
\caption{The flowchart of the route-establishing process in our packet-delivery scheme.} \label{fig:source_transmission}
\end{center}
\vspace{-0.5cm}
\end{figure*}

More specifically, when there is a message delivery requirement from the applications, the source node handles the packet as follows: (1) According to the time list of rejoining and separating, the source determines whether a route exists that makes the message available to the destination at the current moment. (2) If the destination is reachable, the source node calculates the shortest path by utilizing the Dijkstra algorithm based on the adjacency matrix and stores the result in the packet header. Then, the packet-delivery process can start immediately. (3) Otherwise, the source node estimates whether the message is available to the destination before the expiry time. (4) If the destination is reachable before the message expires, the source node calculates the shortest path by using the Dijkstra algorithm with the adjacency matrix as the input, stores the result in the packet header, and waits for the time to transmit. (5) Otherwise, the source node drops the packet proactively and does not attempt to transmit.

\subsection{Data transmission}
\begin{figure}[t]
\centering
\includegraphics[width=0.4\textwidth]{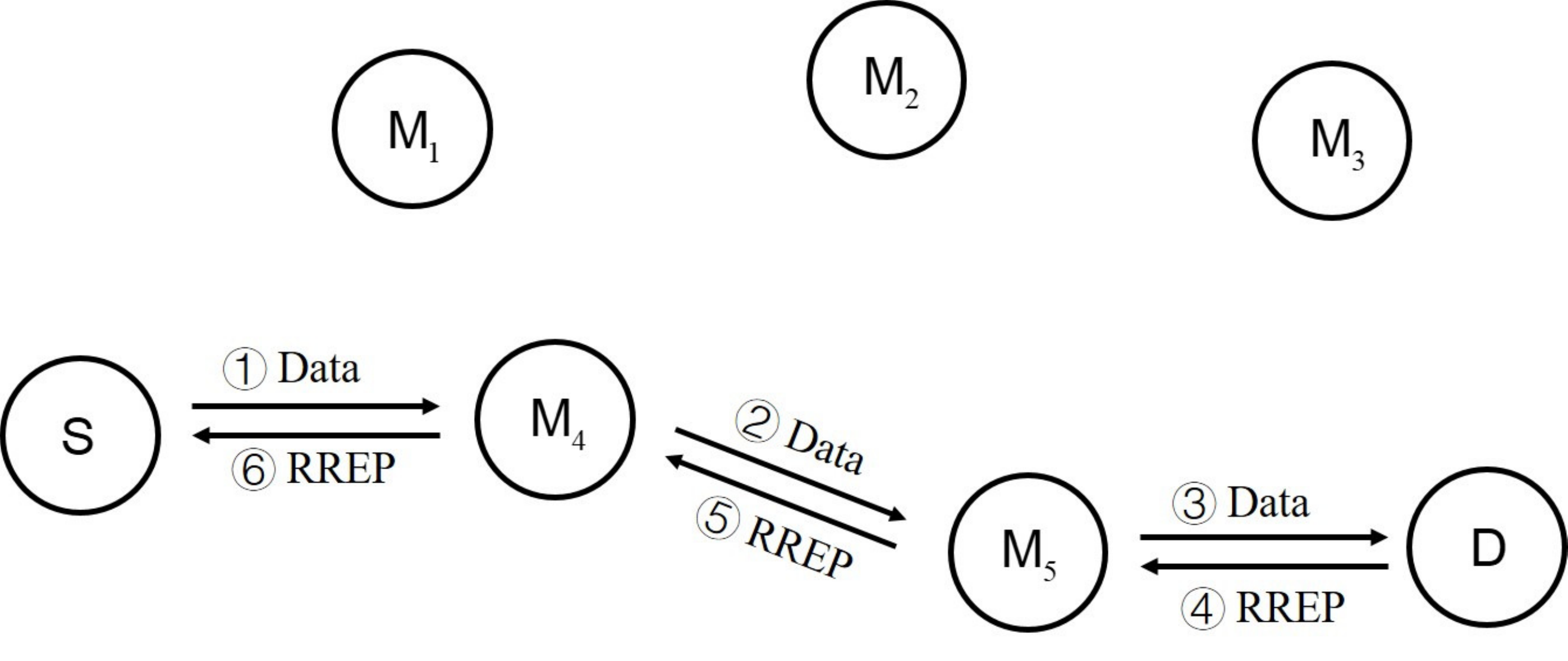}
\caption{The packet-delivery mechanism under the assumption of error-free transmission. $\mathsf S$: source; ${\mathsf M}_1$–${\mathsf M}_5$: intermediate nodes; $\mathsf D$: destination.}
\label{fig:Data_transmission}
\end{figure}

The source node generates the route, while the intermediate nodes only perform data forwarding and feedback. More specifically, when any node receives a Data packet, it judges whether the destination is itself or not. If this receiving node is indeed the destination, it feeds back an RREP packet to the source node, which confirms the successful delivery of the Data packet. Otherwise, this receiving node forwards the Data packet to the appropriate next-hop node, which is determined by the route information stored in the packet header. Assuming error-free transmission, we use an example to illustrate the process of sending the Data and RREP packets in Fig. \ref{fig:Data_transmission}, where each circle represents a node. Specifically, the source $\mathsf S$ sends out a Data packet with the route information stored in the packet header. Upon receiving the Data packet from the source $\mathsf S$, the intermediate node ${\mathsf M}_4$ that is on the calculated route forwards the Data packet to the next intermediate node ${\mathsf M}_5$ on the route. Finally, when the destination $\mathsf D$ receives the Data packet, it feeds back an RREP packet by node relaying, according to the reverse route information stored in the packet header, to the source $\mathsf S$.

\subsection{Route recovery}
\begin{figure}[t]
\centering
\subfloat[]{\includegraphics[width=0.4\textwidth]{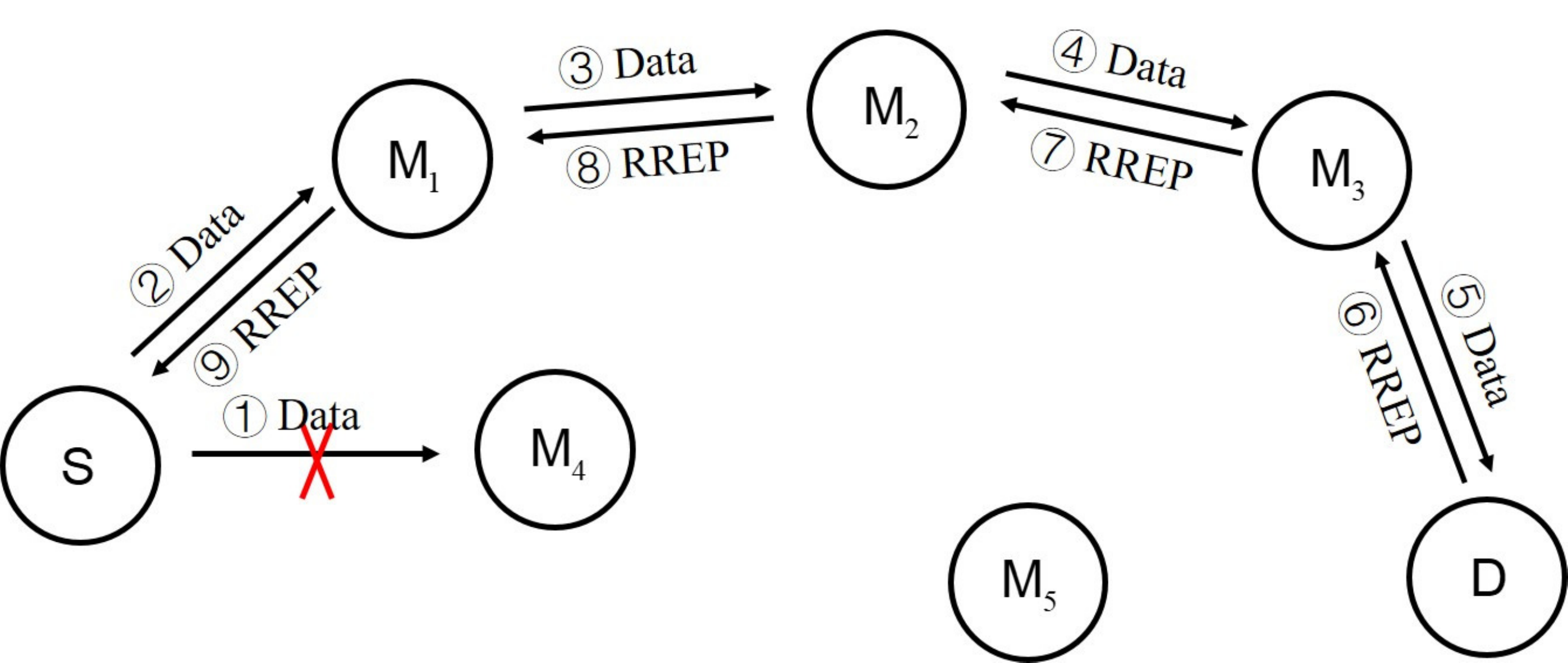}}
\hfil
\subfloat[]{\includegraphics[width=0.4\textwidth]{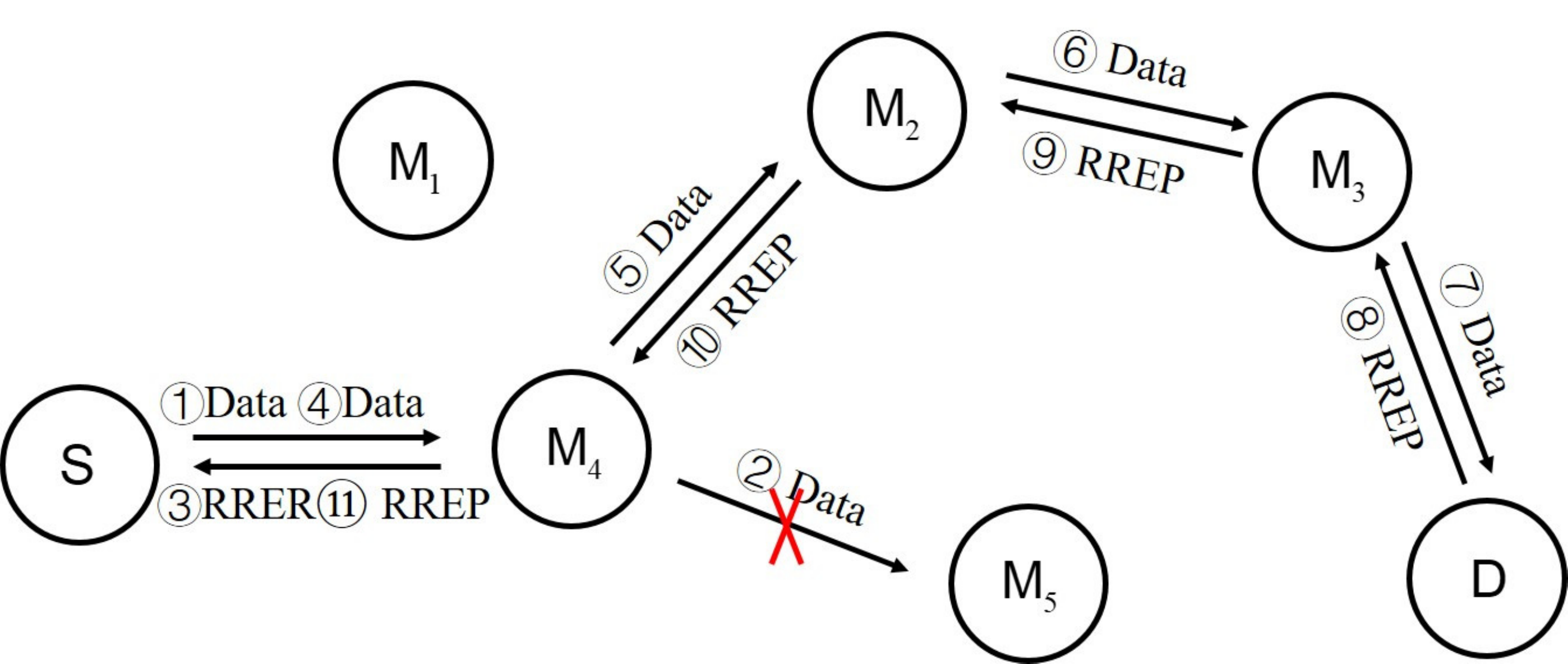}}
\caption{The Data transmission mechanism. (a) Next-hop failure scenario; (b) other-hop failure scenario.}
\label{fig_sim}
\end{figure}

To guarantee highly reliable delivery of the packets across the network, the source node must support retransmission mechanisms in the following two scenarios: next-hop failure and other-hop failure, as shown in Fig.  \ref{fig_sim}. In Fig. 10(a), due to the failure of transmission from the source to the next-hop node, the source will not receive any feedback during the time period that is set for the current environment in advance. Hence, the source must recalculate a route that bypasses the next-hop node where the transmission failure occurred, and then send the Data packet again. In Fig. 10(b), although the next-hop node immediately adjacent to the source node successfully received the Data packet forwarded by the source, one of the other intermediate nodes (excluding the destination) may experience transmission failure. As a result, the source will never receive the RREP packet from the destination. Instead, the destination sends back the RRER packet to the source, thereby acknowledging that there is a transmission failure taking place on one of the links between the intermediate nodes. Then, the source recalculates a route that bypasses the particular intermediate node that experienced the transmission failure. However, if the transmission failure happens on the last hop to the destination, the destination node is not reachable. In this case, the retransmission mechanism will not be invoked.

\section{Simulation results and discussions}\label{sec:simulation}
In this section, we present extensive performance evaluation results of our proposed CPR-TD protocol in the context of an MO-FANET. Our protocol is implemented on an ns-3 network simulator. For the sake of clarity, Table \ref{tab:parameter} summarizes the parameters used in our simulations.

\begin{table}[ht]
\begin{spacing}{1}
\center
\small
\caption{Parameter setting}
\begin{tabular}{@{}ll@{}}
\toprule
Parameter                  & Value                                  \\ \midrule
Network simulator          & ns-3 (release ns-3.30)                      \\
Number of flight vehicles  & 36, 100, and 196                            \\
Speed of flight vehicles   & 250 m/s                                 \\
Radio technology           & 802.11b                                \\
frequency band             & 2.4 GHz                                \\
Transport protocol         & UDP                                    \\
Data rate                  & 11 Mbps                                 \\
PHY protocol               & DSSS                                   \\
MAC protocol               & CSMA/CA                                \\
Packet size                & 1000 bytes                             \\
Packet generating interval                   & 0.1 s                              \\
Traffic model              & CBR                                    \\
Delay model                & Constant speed propagation				\\
Loss model                 & Friis propagation          		    \\
Transmission power         & 16.0206 dbm                             \\
Energy detection threshold & -96 dbm                                 \\
Mobility model             & Adaptive flight formation                        \\
Simulation time            & 100 s                                   \\ \bottomrule
\end{tabular}
\label{tab:parameter}
\begin{tablenotes}
\footnotesize
UDP: user datagram protocol; PHY: physical layer; DSSS: direct sequence spread
spectrum; MAC: medium access control; CSMA/CA: carrier sense multiple
access/collision avoidance; CBR: constant bit rate.
\end{tablenotes}
\end{spacing} 
\end{table}

\subsection{Trajectory configuration}\label{subsec:trajectory}
In order to evaluate the performance of our protocol in the simulated network model, the dynamic trajectories of the MO-FANET during a given operation period are divided into five phases, as shown in Table \ref{tab:Trajectory_Parameter}. Each phase represents a distinct state of motion, as shown in Fig. \ref{fig:Flight_formation} and explained below.

\begin{table}[htbp]
\begin{spacing}{1.0}
\small
  \centering
  \caption{Trajectory configuration.}
    \begin{tabular}{@{}lll@{}}
    \toprule
    Phase & Time periods & Flight scenarios \\	\midrule
    1     & [0, 30.1) & Transition to diversion by maneuver \\
    2     & [30.1, 37.7) & Transition to diversion by maneuver  \\
    3     & [37.7, 60.1) & Fly in diversion \\
    4     & [60.1, 62.8) & Transition to rendezvous in batches \\
    5     & [62.8, 100] & Fly in rendezvous and adjust topology \\ \bottomrule
    \end{tabular}%
  \label{tab:Trajectory_Parameter}%
  \end{spacing} 
\end{table}%

\begin{itemize}
\item In phase 1, all the UAVs fly in close formation and carry out a uniform rectilinear motion with a relatively stable topology.
\item In phase 2, in order to avoid emerging obstacles, the single formation transitions to several sub-formations by maneuvering. In other words, the network shifts from cohesion to diversion. More specifically, each UAV group first establishes an obstacle-avoidance maneuvering strategy within their individual airborne control systems; the individual UAVs then start to climb, descend, and turn left/right, respectively. In this process, the UAVs in each group can still maintain communication with each other, but UAVs from different groups may gradually lose contact with each other.
\item  In phase 3, each group is out of the communication range of other groups, while the formation is well kept among the UAVs in each group. As a result, the packet can only be delivered inside each group.
\item In phase 4, the sub-formations begin to rendezvous at the scheduled time and gradually become close to each other. Meanwhile, communications between the sub-formations are gradually reestablished.
\item  In phase 5, the sub-formations have completed the assembly from the communications perspective; they then adjust their motion to fly in formation again.
\end{itemize}

\subsection{Performance metrics}
For the sake of performance evaluation, we consider two realistic and typical operating scenarios. In the first scenario, all the UAVs are in good condition and fly according to the schedule. In the second scenario, one of the UAVs within each group is assumed to function improperly. We define several performance metrics as follows:
\begin{itemize}
\item[(1)]  \textbf{PDR.} This is defined as the ratio of the total number of received packets,$N_{\textrm{received}}$, to the total number of packets sent from the source to the destination, $N_{\textrm{sent}}$; that is, we have
\begin{equation}
{\mathsf {PDR}} = \frac{N_{\textrm{received}}} {N_{\textrm{sent}}},
\end{equation}
which reflects the transmission reliability of the network using a particular routing protocol.
\item[(2)] \textbf{OE.} This is defined as the ratio of the size of all the Data packets received (i.e., the payload, $N_{\textrm{payload}}$) to the size of all the Control packets sent (i.e., the overhead, $N_{\textrm{overhead}}$); that is, we have
\begin{equation}
\mathsf {OE} = \frac{N_{\textrm{payload}}} {N_{\textrm{overhead}}} = \frac{{\sum {{\mathsf {DP}}_{\textrm{size},i}}}} {\sum {\mathsf {CP}}_{\textrm{size},j}},
\end{equation}
where ${\mathsf {DP}}_{\textrm{size},i}$ is the size of the $i$th Data packet; ${\mathsf {CP}}_{\textrm{size},i}$ is the size of the $j$th Control packet; and $\mathsf {OE}$ characterizes how efficient the overhead is used in a routing protocol invoked by the network. The reception of more Data packets at the cost of less Control packets indicates that the routing protocol is working at higher efficiency.
\item[(3)]  \textbf{Average end-to-end latency.} This is defined as the average time for delivering a packet across the network from a source to a destination; that is, we have
\begin{equation}
{\overline T}_{\text{\textrm{E2E}}} = \frac{\sum ({T_{\textrm{received}}} - {T_{\textrm{sent}}} ) } {N_{\textrm{received}}},
\end{equation}
where $T_{\textrm{received}}$ and $T_{\textrm{sent}}$ represent the time instant of receiving a packet at the destination and sending a packet from the source, respectively. ${\overline T}_{\textrm{E2E}}$ is a key factor in determining whether the network meets the latency requirements of transmission.

\item[(4)] \textbf{Network jitter.} This is defined as the standard deviation of the end-to-end latency $T_{\textrm{E2E}} = T_\textrm{received} - T_\textrm{{sent}}$; that is, we have

\begin{equation}
{T_{\textrm{jitter}}} = \frac{{\sqrt {\sum {{{(T_{\textrm{E2E},i}} -  {\overline T}_{\textrm{E2E}})}^2}} } }  {{N_{\textrm{received}} - 1}},
\end{equation}
which characterizes the network stability.
\end{itemize}

\subsection{Results and discussions}
By developing an ns-3-based network simulator with the parameters specified in Table \ref{tab:parameter}, Table \ref{tab:Trajectory_Parameter}, the proposed CPR-TD protocol is comprehensively compared with five state-of-the-art routing protocols: the AODV, the destination-sequenced distance vector (DSDV), the DSR, the OLSR, and the greedy perimeter stateless routing (GPSR) protocols.

AODV is a topology-based on-demand routing protocol. Each node in AODV maintains a routing table by means of the control packets Hello, RREQ, RRER, and RREP [24]. The routing table contains one route entry for each known destination node in the network, and a route from the origin to the destination is discovered only when routes are needed. DSDV is a proactive routing protocol originating from the idea of the Bellman–Ford algorithm. Each node periodically sends its routing table to the neighbor nodes, recomputes the shortest distance, and updates the table [25]. In DSR, the RREP and RREQ packets are also used for route discovery, and the routes are stored in the packet header when transmission begins [26]. For a particular node, by declaring it as an MPR selector of each neighbor node, OLSR reduces the size of control packets [24]. In GPSR, the global route is assigned for each node according to the geographic positions of its neighbors and the transmission endpoints, using a greedy algorithm [27].

\begin{figure}[t]
\centering
\includegraphics[width=0.5\textwidth]{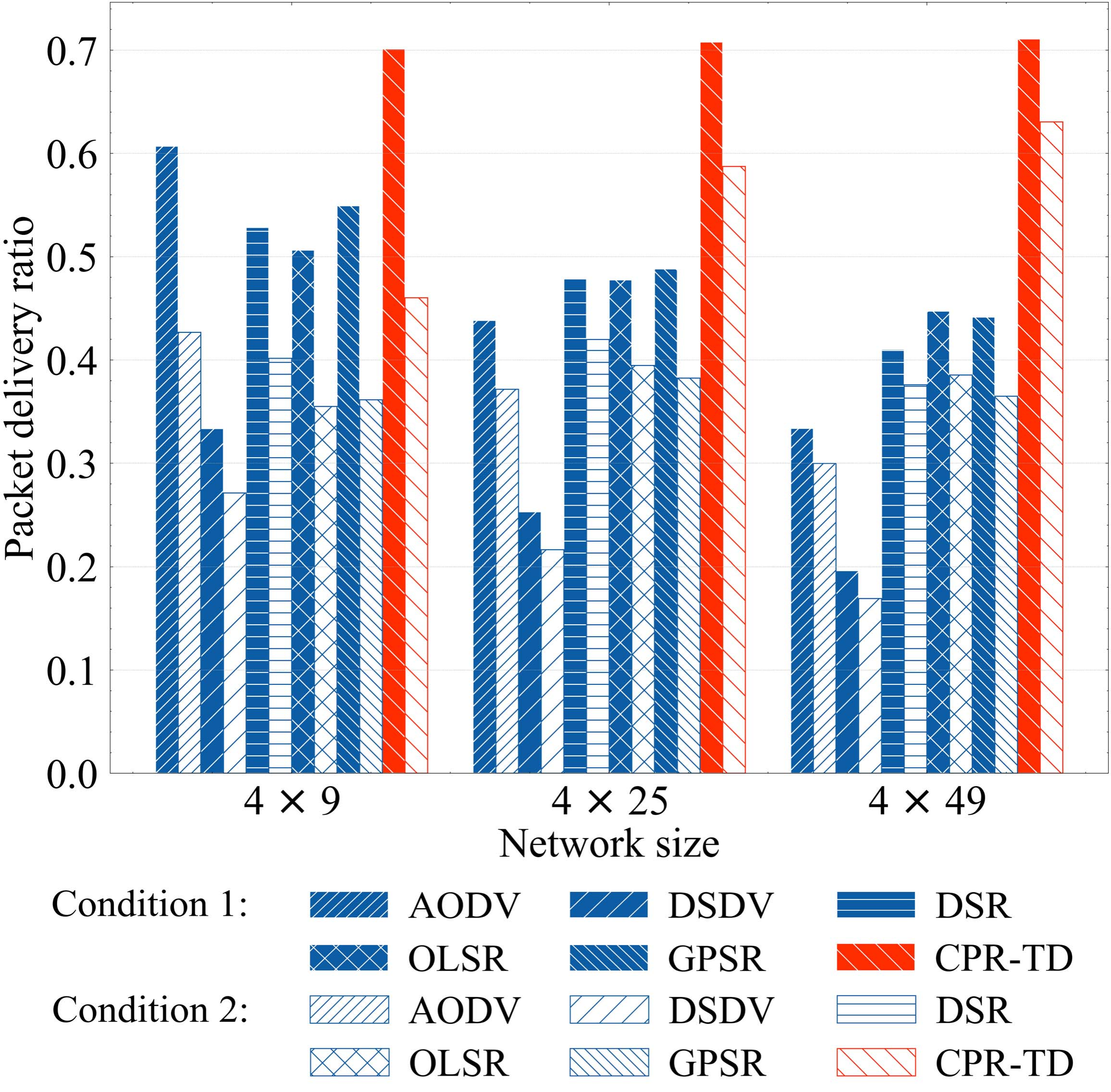}
\caption{PDR comparison of different routing protocols under different values of network size and different node-failure conditions.}
\label{fig:pdr_totoal}
\end{figure}

Four performance metrics—namely, the PDR, the OE, the average end-to-end latency, and the network jitter—are considered under different values of network size (i.e., the number of nodes is set to $4\times 9 = 36$, $4\times 25 =100$ and $4\times 49 = 196$, with “4” representing the number of sub-formations in the system), different phases of motion, and different node-failure conditions. The results of each routing protocol under consideration are obtained through 100 Monte Carlo simulation experiments; in each Monte Carlo experiment, a total of 990 Data packets are sent from randomly selected source nodes to randomly selected destination nodes during the 100 s simulation period. It is notable that the simulator does not work at the time instants of 0 s and 100 s due to the inherent mechanism of ns-3, and the packet-generating interval is 0.1 s. Thus, the first Data packet is set to be sent at the time instant of 1.0 s, and the last Data packet transmission is finished at the time instant of 99.9 s.

In Fig. \ref{fig:pdr_totoal}, we compare the PDR of different routing protocols under different values of network size. Meanwhile, we consider two node-failure conditions. Condition 1 represents a scenario in which all the nodes are working well, whereas in Condition 2, the central node in each group is assumed to misfunction. It can be seen that the proposed CPR-TD protocol achieves the highest PDR in comparison with the other benchmarking protocols, which is true even under the node failure of Condition 2. This is because the proposed CPR-TD conveniently benefits from the motion prediction obtained by exploiting the trajectory dynamics. Thus, it does not need the route search and maintenance process. As a result, there is a lower probability of medium-access conflict being incurred by the route-discovery operation.

\begin{figure*}[t]
\begin{center}
\includegraphics[angle = 0,width = 0.98\textwidth]{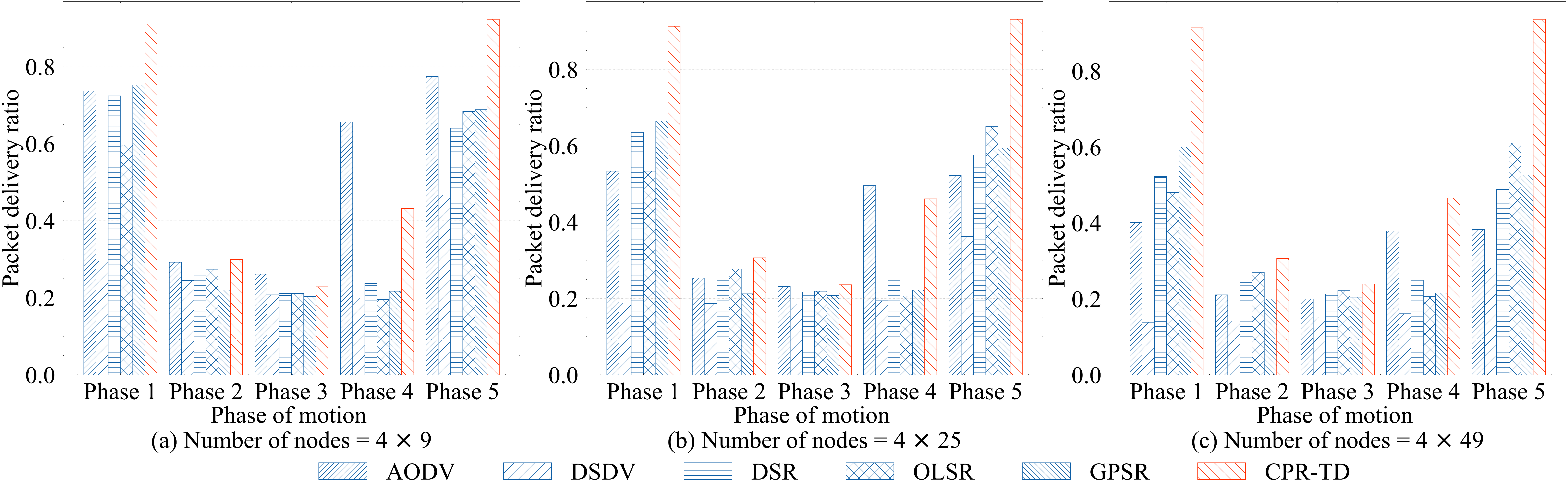}
\caption{PDR comparison of different routing protocols in different phases of the flying process and under different values of network size. (a) number of nodes is $4\times 9 $; (b) number of nodes is $4\times 25 $; (c) number of nodes is $4\times 49 $.}
\label{fig:pdr_phase}
\end{center}
\vspace{-0.5cm}
\end{figure*}

\begin{figure}[t]
\centering
\includegraphics[width=0.45\textwidth]{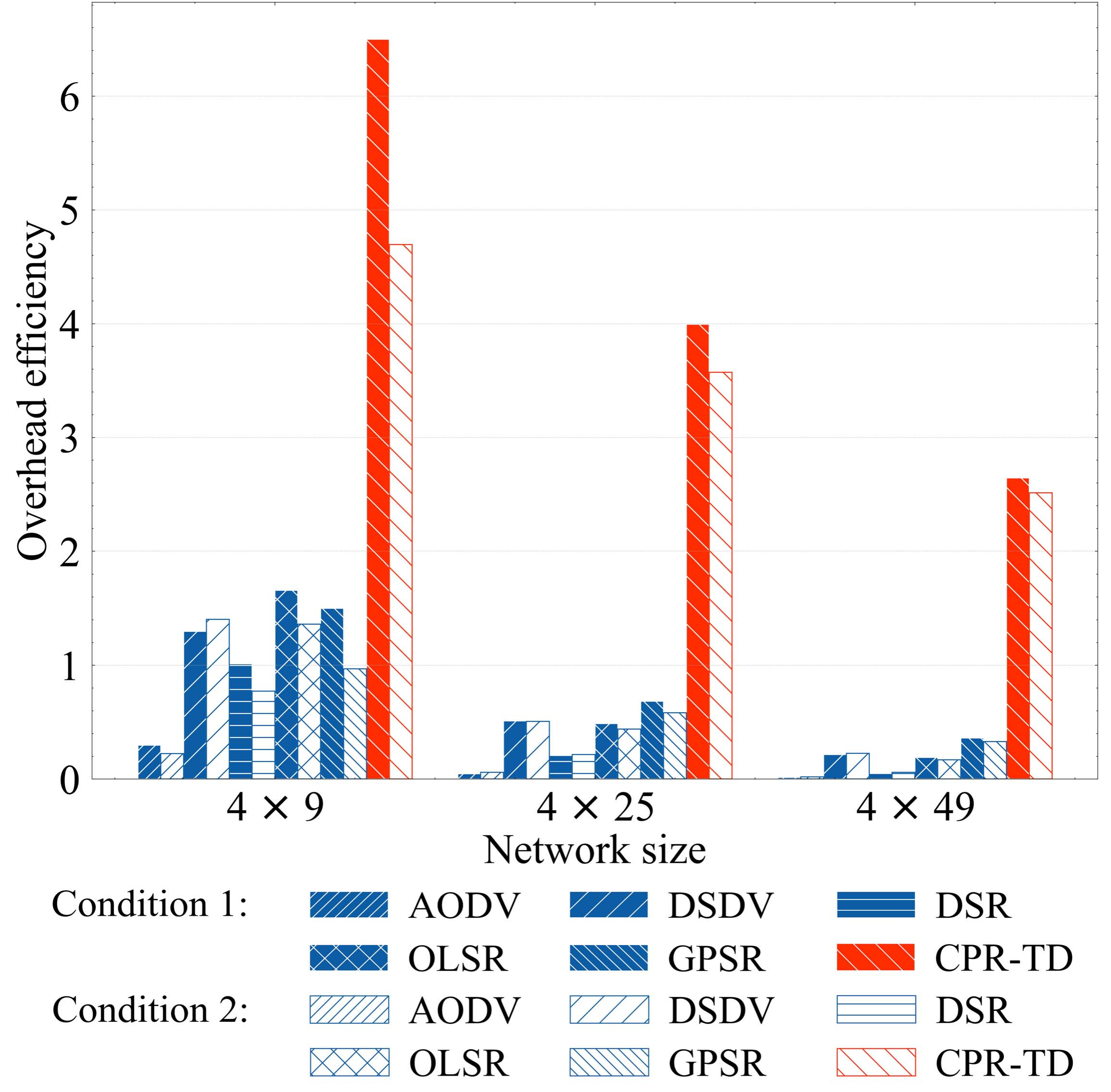}
\caption{OE comparison of different routing protocols under different values of network size and different node-failure conditions.}
\label{fig:overhead_totoal}
\end{figure}

\begin{figure*}[htbp]
\begin{center}
\includegraphics[angle = 0,width = 0.98\textwidth]{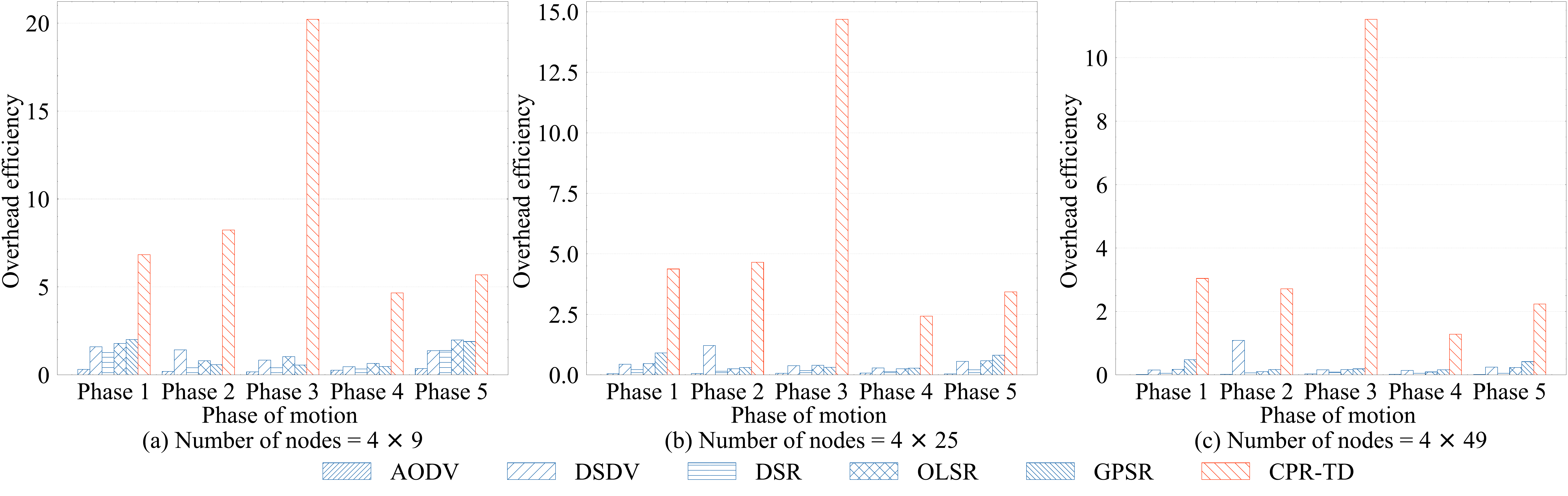}
\caption{OE comparison of different routing protocols in different phases of the flying process and under different values of network size. (a) number of nodes is $4\times 9 $; (b) number of nodes is $4\times 25$; (c) number of nodes is $4\times 49 $.}
\label{fig:overhead_phase}
\end{center}
\vspace{-0.5cm}
\end{figure*}
Under both Condition 1 and Condition 2, the DSDV protocol exhibits the worst PDR among all the considered protocols. This is because DSDV is a proactive routing protocol, which requires every node in the network to send Control packets periodically, and thus has the highest probability of medium-access conflict, regardless of whether there is a data transmission request or not. In contrast, all the other protocols are relatively cautious in calling for the nodes in the network to maintain routing tables, thus incurring less overhead. In addition, the PDR performance of the AODV protocol degrades the most as the network size grows, while the DSR, OLSR, and GPSR protocols achieve somewhat similar PDR performance. Finally, we see that the impact of node failure is reduced with an increase in network size for all the protocols. This is a natural result, since the density of node failure becomes smaller when the network size becomes larger.

A PDR comparison of different routing protocols in different phases of the flying process and under different values of network size is shown in Fig. 12. Without loss of generality, we consider a scenario in which all the nodes function well. Different from the observations obtained from Figs. 12(a) and (b), it can be seen in Fig. 12(c) that the proposed CPR-TD achieves the best performance in all the five phases specified in Table 2, when the network size is sufficiently large. Although this advantage is degraded slightly when the network size is small, our CPR-TD protocol still outperforms the other benchmarking protocols in Phase 1, Phase 2, and Phase 5, as shown in Figs. 12(a) and (b), and only the AODV protocol exhibits marginally higher PDR than our CPR-TD in Phase 3 and Phase 4. Moreover, it can be seen that the PDR performance of all the protocols is degraded dramatically in Phase 2, Phase 3, and Phase 4 compared with that of Phase 1 and Phase 5. This is because the connectivity of the network is high in Phase 1 and Phase 5, while in Phase 2 and Phase 3, the nodes are in the process of transitioning from a single formation to several sub-formations by maneuvering, and in the state of well-separated multiple sub-formations, respectively. Consequently, the network connectivity is weak in Phase 2 and Phase 3. It is notable that AODV performs well in Phase 4, where the nodes are in the process of transitioning from being separated to rejoining a single formation. This is because AODV has a highly effective route-discovery mechanism to find opportunities to send messages. However, its superiority declines as the network size increases.

Fig. 13 shows the OE versus network size performance of different routing protocols. Again, two node-failure conditions are considered, as in Fig. 11. It is clear that our CPR-TD protocol has the highest OE under both Condition 1 and Condition 2, which means that it delivers the largest amount of data per byte of overhead. In contrast, all the benchmarking protocols exhibit a significantly lower OE. This is because these protocols commonly invoke a route-discovery mechanism, which assumes that the random motion of the nodes cannot be predicted in an ordinary MANET. However, in the MO-FANET being considered here, the fact is that the UAVs must be under control. Thus, the motion characteristics of the nodes can be acquired by calculating the trajectory dynamics, and can then serve as a particular type of prior information for a tailored routing protocol. In addition, under the node failure in Condition 2, the overhead of our CPR-TD protocol increases moderately, which is consistent with our expectation that some overhead will be paid for route recovery under this condition. It is also notable that, among the benchmarking protocols, DSDV, OLSR, and GPSR have a higher OE than AODV and DSR. This finding points to the relative amount of overhead they incurred in their different route-discovery mechanisms. Finally, it is notable that, in theory, our CPR-TD protocol can also report the nodes’ information to the trajectory planning application in return, thus resulting in a cyber–physical close-loop design.

The OE versus network size performance of the different routing protocols in different phases of the flying process is demonstrated in Fig. 14, under the assumption that all the nodes are functioning well. It can be seen that, in all five phases and all network size values considered, our CPR-TD protocol achieves the highest OE, while AODV exhibits the lowest OE. This observation is consistent with that obtained from Fig. 13. Again, this is because the route-discovery mechanism in AODV causes the largest overhead in each phase, while the opposite is true for our CPR-TD.

\begin{figure}[ht]
\centering
\includegraphics[width=0.45\textwidth]{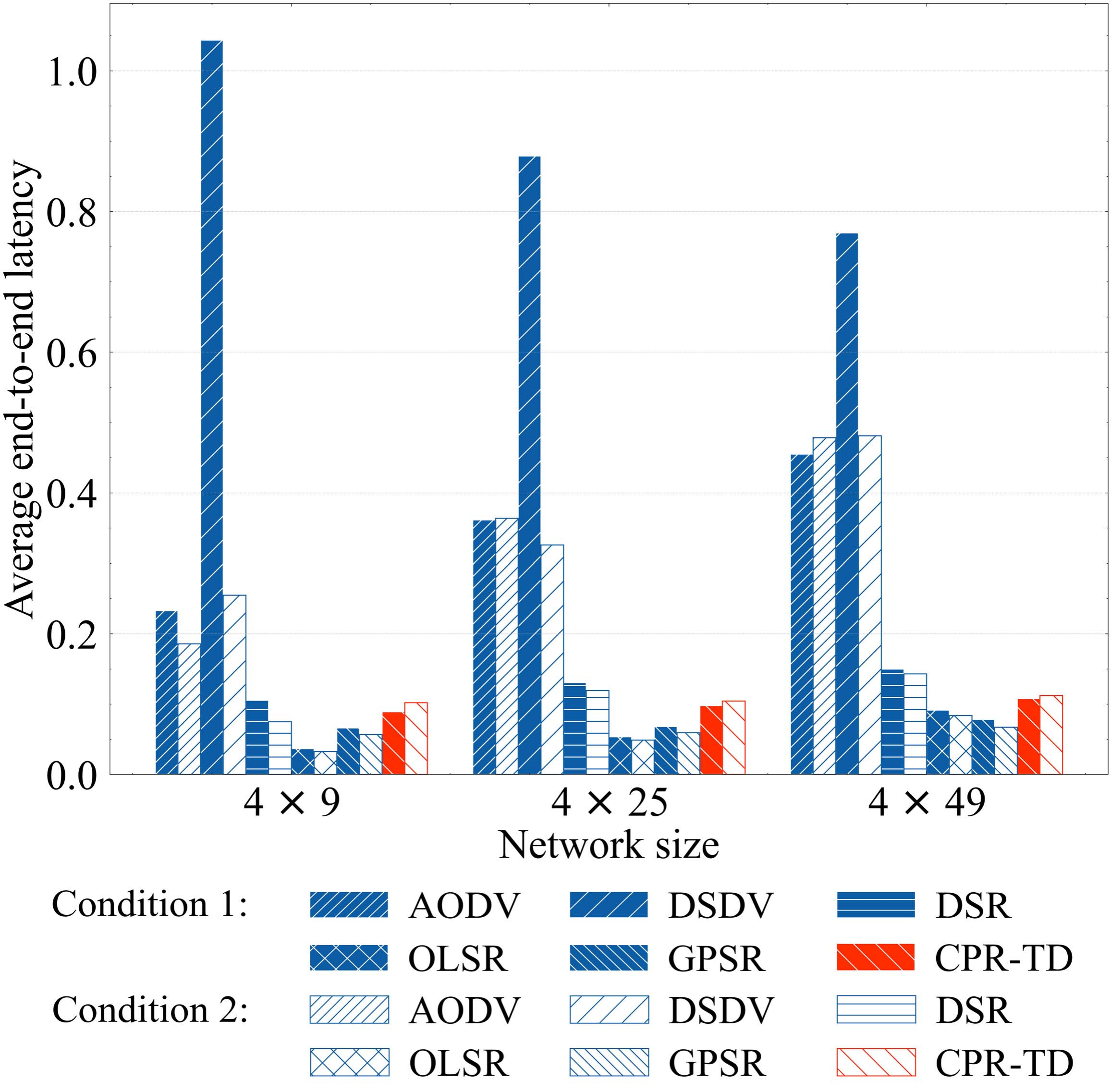}
\caption{A comparison of average end-to-end latency for different routing protocols under different values of network size and different node-failure conditions.}
\label{fig:delay_totoal}
\end{figure}

In Fig. 15, we compare the average end-to-end latency of different routing protocols under different values of network size and different node-failure conditions. In Condition 1, it can be seen that, although the latency of DSDV is reduced with an increase in network size, it remains the largest among those of all the routing protocols under consideration. This is because only DSDV requires all the nodes in the network to send Control packets periodically and, with an increase in network size, its PDR remains the worst (Fig. 11). When the network size is large, for DSDV, only the nodes that are close to each other are involved in the packet-delivery process, which results in decreased latency. Therefore, DSDV does not fit time-sensitive application scenarios. For the rest of the routing protocols, the latency generally becomes larger as the network size increases, although the extent of the latency increase is different for specific protocols. More specifically, AODV and DSR exhibit the second- and third-largest latency, respectively; OLSR exhibits the smallest latency when the network size is small, but its latency increases significantly when the network size becomes large; and GPSR and our CPR-TD show relatively stable latency performance with an increase in network size. Our CPR-TD has a reasonably moderate latency performance, because the source still needs to calculate the shortest path, which takes time, and the header of the packet needs to be extracted and processed during the message-forwarding process. Overall, our CPR-TD protocol performs well in large-scale networks. Finally, under Condition 2, it can be seen that the node failure has a marginal impact on the average end-to-end latency performance of the routing protocols, with the exception of DSDV.
\begin{figure*}[htbp]
\begin{center}
\includegraphics[angle = 0,width = 0.98\textwidth]{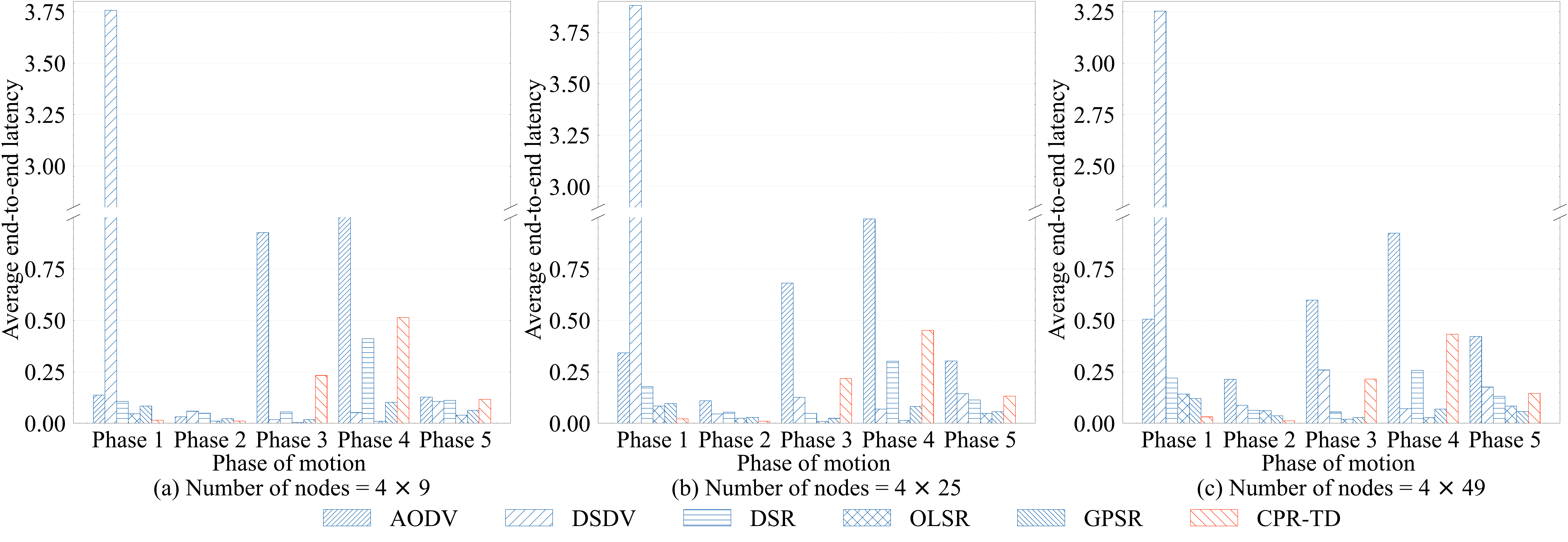}
\caption{A comparison of average end-to-end latency for different routing protocols in different phases of the flying process and under different values of network size. (a) number of nodes is $4\times 9$; (b) number of nodes is $4\times 25$; (c) number of nodes is $4\times 49$.}
 \label{fig:delay_phase}
\end{center}
\vspace{-0.5cm}
\end{figure*}

Assuming Condition 1 and different network sizes, Fig. 16 shows the average end-to-end latency performance of different routing protocols in different phases of the flying process. It can be seen that, under all three values of network size, DSDV exhibits dramatically larger average end-to-end latency than the other protocols in Phase 1. However, from Phase 2 to Phase 5, when the network size is sufficiently large (e.g.,$4\times 25$ and $4\times 49$), AODV shows the largest latency. This is because DSDV proactively maintains routing tables in Phase 1 with high network connectivity, thereby causing more medium-access conflicts and more nodes to be involved in the transmission route, eventually resulting in a huge average end-to-end latency. But when the network connectivity is reduced (e.g., from Phase 2 to Phase 5), the route-discovery mechanism of AODV weighs most in terms of latency. Our CPR-TD protocol has a relatively large latency in Phase 4, since the source must wait for the time of rejoining to send out the Data packets, if they are available, to the destination before they expire.

Finally, Fig. 17 demonstrates the network jitter performance of different routing protocols under different values of network size and different node-failure conditions. It can be seen that the network jitter of our CPR-TD, although not the lowest, remains competitive. In terms of this metric, the relative pros and cons of different protocols, as well as the corresponding reasons, are similar to those revealed for Fig. 15. Taking Fig. 11, Fig. 13, Fig. 15, Fig. 17 into consideration, we can conclude that our CPR-TD achieves the most attractive performance tradeoff among the PDR, OE, average end-to-end latency, and network jitter.
\begin{figure}[ht]
\centering
\includegraphics[width=0.45\textwidth]{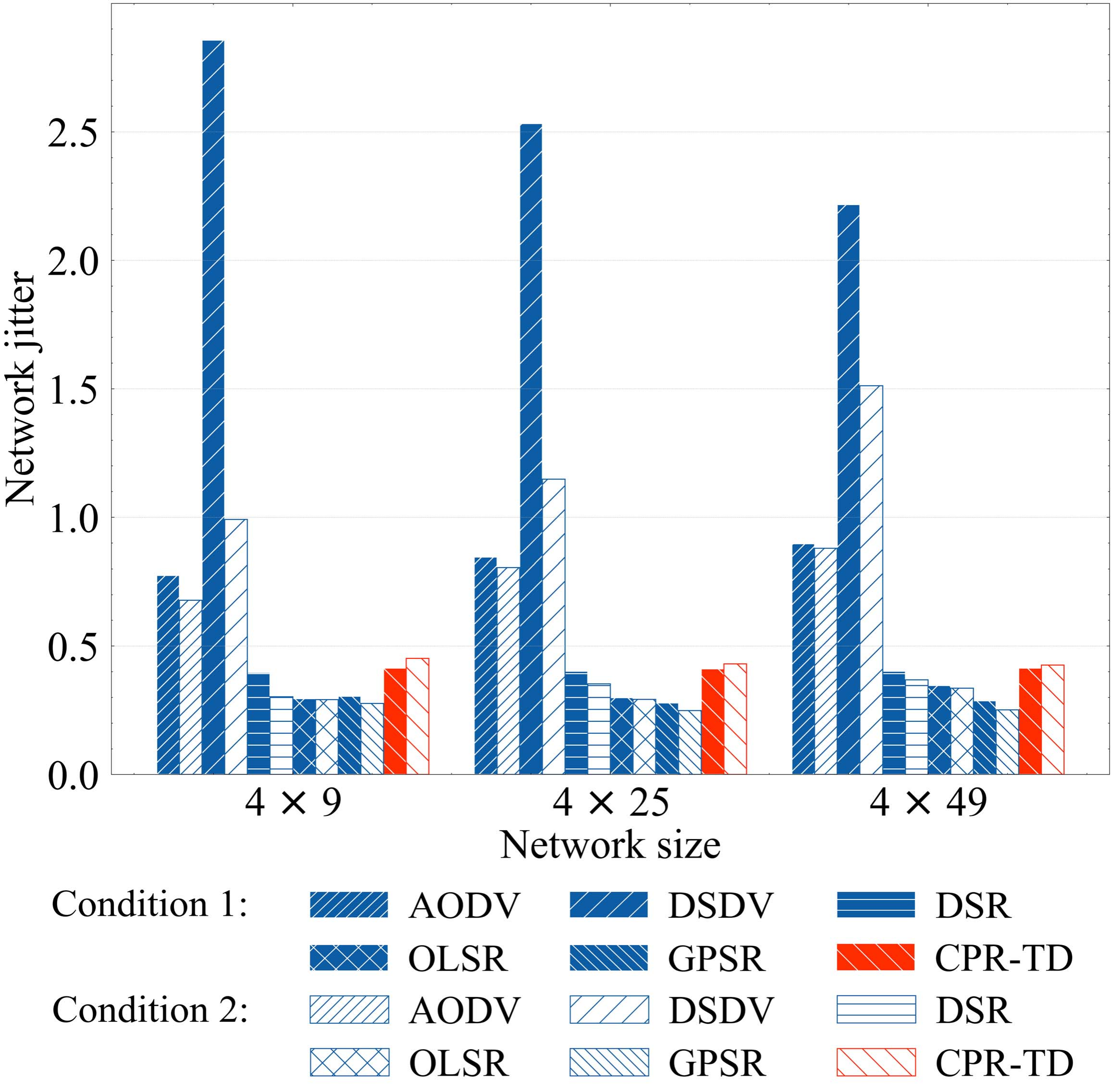}
\caption{Comparison of network jitter.}
\label{fig:jitter_totoal}
\end{figure}

\section{Conclusion}
In this paper, we proposed a CPR-TD protocol for an MO-FANET, which holds potential for diverse applications in both civilian and defense industries. A challenging scenario was considered in which multiple UAVs fly collaboratively in different formations from one place to another in order to execute a certain task. The flying process of the MO-FANET was modeled as five phases, which were described by multiple coordinate frames and sophisticated maneuver actions. Benefiting from the cross-disciplinary integration of wireless networks and trajectory dynamics, our CPR-TD protocol can be implemented with the aid of the individual ATP system and a cross-layer protocol stack. The performance of the proposed CPR-TD protocol was compared with that of five representative routing protocols used in FANETs. Extensive simulations based on ns-3 while assuming realistic network configurations demonstrated that our CPR-TD protocol not only achieved the highest PDR performance, but also attained the highest OE. Moreover, in most of the considered cases, it exhibited a lower average end-to-end latency than the benchmarking protocols, as well as a reasonably low and stable network jitter.

In our future work, we intend to conduct the following in-depth research: (1) Investigate how to access and use the information from other layers to serve the routing protocol, in order to improve the overall performance of the network; (2) investigate how to utilize distributed algorithms to accelerate the computation and realize lower processing latency, in order to support real-time decision-making; and (3) apply the protocol in a real MO-FANET to test its performance and conduct further optimizations.

\section*{Acknowledgment}

This work is financially supported by the Beijing Municipal Natural Science Foundation (L202012), the Open Research Project of the State Key Laboratory of Media Convergence and Communication, Communication University of China (SKLMCC2020KF008), and the Fundamental Research Funds for the Central Universities (2020RC05).

The authors would like to thank Professor Ping Zhang (Member of the Chinese Academy of Engineering, Beijing University of Posts and Telecommunications) and Professor Quan Yu (Member of the Chinese Academy of Engineering, Peng Cheng Laboratory) for their insightful comments and suggestions.

\section*{Compliance with ethics guidelines}
Die Hu, Shaoshi Yang, Min Gong, Zhiyong Feng, and Xuejun Zhu declare that they have no conflict of interest or financial conflicts to
disclose.

\section{Reference}

\begin{itemize}
\item[1.]
Maza I, Caballero F, Capit$\acute{a}$n J, Mart$\acute{i}$nez-de-Dios JR, Ollero A. Experimental results in multi-UAV coordination for disaster management and civil security applications. J Intell Robot Syst 2011; 61(1–4):563–85.
\item [2.]
Vollgger SA, Cruden AR. Mapping folds and fractures in basement and cover rocks using UAV photogrammetry, Cape Liptrap and Cape Paterson, Victoria, Australia. J Struct Geol 2016; 85:168–87.
\item[3.]
Meng XY, Wang W, Leong B. SkyStitch: A cooperative multi-UAV-based realtime video surveillance system with stitching. In: Proceedings of 23rd ACM International Conference on Multimedia; 2015 Oct 26–30;  Brisbane, QSD, Australia. ACM;  2015. p. 261–70.
\item[4.]
Cheng X, Lyu F, Quan W, Zhou C, He H, Shi W, et al. Space/aerial-assisted computing offloading for IoT applications: a learning-based approach. IEEE J Sel Area Commun 2019; 37(5):1117–29.
\item[5.]
Zhao E, Chao T, Wang S, Yang M. Finite-time formation control for multiple flight vehicles with accurate linearization model. Aerosp Sci Technol 2017; 71: 90–8.
\item[6.]
Quan W, Cheng N, Qin M, Zhang H, Chan HA, Shen X. Adaptive transmission control for software defined vehicular networks. IEEE Wirel Commun Lett 2019; 8(3):653–6.
\item[7.]
Oubbati OS, Lakas A, Zhou F, G$ddot(u)$nes M, Yagoubi MB. A survey on position-based routing protocols for flying ad hoc networks (FANETs). Veh Commun 2017; 10: 29–56.
\item[8.]
Alshbatat AI, Dong L. Cross layer design for mobile ad-hoc unmanned aerial vehicle communication networks. In: Proceedings of 2010 International Conference on Networking, Sensing and Control (ICNSC);  2010 Apr 10–12;  Chicago, IL, USA. New York City: IEEE;  2010. p. 331–6.
\item[9.]
Paul AB, Nandi S. Modified optimized link state routing (M-OLSR) for wireless mesh networks. In: Proceedings of 2018 International Conference on Information Technology;  2008 Dec 17–20;  Bhubaneswar, India. New York: IEEE;  2008. p. 147–52.
\item[10.]
Park SY, Shin CS, Jeong D, Lee H. DroneNetX: Network reconstruction through connectivity probing and relay deployment by multiple UAVs in ad hoc networks. IEEE Trans Veh Technol 2018; 67(11):11192–207.
\item[11.]
Shirani R, St-Hilaire M, Kunz T, Zhou Y, Li J, Lamont L. On the delay of reactivegreedy- reactive routing in unmanned aeronautical ad-hoc networks. Procedia Comput Sci 2012; 10:535–42.
\item[12.]
Perkins CE, Royer EM. In: Ad-hoc on-demand distance vector routing. New Orleans, LA, USA. New York City: IEEE;  1999. p. 90–100.
\item[13.]
Camp T, Boleng J, Davies V. A survey of mobility models for ad hoc network research. Wirel Commun Mob Comput 2002; 2(5):483–502.
\item[14.]
Cheng N, Quan W, Shi W, Wu H, Ye Q, Zhou H, et al. A comprehensive simulation platform for space–air–ground integrated network. IEEE Wirel Commun 2020; 27(1):178–85.
\item[15.]
S$\acute{a}$nchez M, Manzoni P. ANEJOS: A Java based simulator for ad hoc networks. Future Gener Comput Syst 2001; 17(5):573–83.
\item[16.]
Liang B, Haas ZJ. Predictive distance-based mobility management for multidimensional PCS networks. IEEE/ACM Trans Network 2003; 11(5): 718–32.
\item[17.]
Lin Z, Wang L, Han Z, Fu M. A graph Laplacian approach to coordinate-free formation stabilization for directed networks. IEEE Trans Autom Control 2016; 61(5):1269–80.
\item[18.]
Wang R, Dong X, Li Q, Ren Z. Distributed adaptive formation control for linear swarm systems with time-varying formation and switching topologies. IEEE Access 2016; 4:8995–9004.
\item[19.]
Ning Q, Tao G, Chen B, Lei Y, Yan H, Zhao C. Multi-UAVs trajectory and mission cooperative planning based on the Markov model. Phys Commun 2019; 35: 1–10.
\item[20.]
Dong X, Hua Y, Zhou Y, Ren Z, Zhong Y. Theory and experiment on formationcontainment control of multiple multirotor unmanned aerial vehicle systems. IEEE Trans Autom Sci Eng 2019; 16(1):229–40.
\item[21.]
Zhong D, Zhang H, Chen K. Trajectory generator of SINS based on flight mechanics and control in simulink. In: Proceedings of 2018 Chinese Automation Congress (CAC);  2018 Nov 30–Dec 2;  Xi’an, China. New York City: IEEE;  2018. p. 1638–43.
\item[22.]
 Chen K, Wang X, Liu M, Yu Y, Yan J. Coordinate transformation with application in HWIL simulation. Command Control Simul 2017; 39(2): 118–22. Chinese.
\item[23.]
Butcher JC. The numerical analysis of ordinary differential equations: Runge-Kutta and general linear methods. Chichester: John Wiley $\And$ Sons Ltd;  1987.
\item[24.]
Ferronato JJ, Trentin MAS. Analysis of routing protocols OLSR, AODV and ZRP in real urban vehicular scenario with density variation. IEEE Lat Am Trans 2017; 15(9):1727–34.
\item[25.]
Malik FM, Khattak HA, Almogren A, Bouachir O, Din IU, Altameem A. Performance evaluation of data dissemination protocols for connected autonomous vehicles. IEEE Access 2020; 8:126896–906.
\item[26.]
Bai R, Singhal M. DOA: DSR over AODV routing for mobile ad hoc networks. IEEE Trans Mobile Comput 2006; 5(10):1403–16.
\item[27.]
Chen Q, Kanhere SS, Hassan M. Adaptive position update for geographic routing in mobile ad hoc networks. IEEE Trans Mobile Comput 2013; 12(3): 489–501. 
\end{itemize}

\ifCLASSOPTIONcaptionsoff
  \newpage
\fi


%

\end{document}